\documentclass[12pt]{article}
\usepackage{fullpage}
\usepackage{epsf}
\usepackage{epsfig}
\usepackage{amsthm}
\usepackage{fancyhdr} 
\usepackage{amsfonts}
\usepackage{leftidx}
\usepackage{color}
\usepackage{amsmath}
\usepackage{setspace}
\usepackage[numbers]{natbib}
\usepackage{hyperref}
\usepackage{algorithmic}
\usepackage[boxed]{algorithm}
\usepackage[usenames, dvipsnames]{xcolor} 
\usepackage{graphicx}
\usepackage{booktabs}
\usepackage{multirow}

\begin{document}

\newcommand{\bm}[1]{\mbox{\boldmath $#1$}}
\newcommand{\mb}[1]{#1}
\newcommand{\bE}[0]{\mathbb{E}}
\newcommand{\bV}[0]{\mathbb{V}\mathrm{ar}}
\newcommand{\bP}[0]{\mathbb{P}}
\newcommand{\ve}[0]{\varepsilon}
\newcommand{\mN}[0]{\mathcal{N}}
\newcommand{\iidsim}[0]{\stackrel{\mathrm{iid}}{\sim}}
\newcommand{\NA}[0]{{\tt NA}}
\newcommand{\cB}{\mathcal{B}}
\newcommand{\R}{\mathbb{R}}
\newcommand{\Rp}{\R_+}


\title{\vspace{-1cm}Synthesizing simulation and field data of solar irradiance}
\author{Furong Sun\thanks{Corresponding author: Department of Statistics, Virginia Tech,
Hutcheson Hall, 250 Drillfield Drive
Blacksburg, VA 24061, USA ;
\href{mailto:furongs@vt.edu}{\tt furongs@vt.edu}}
\and Robert B.~Gramacy\thanks{Department of Statistics, Virginia Tech}
\and Benjamin Haaland\thanks{Population Health Sciences, University of Utah} 
\and Siyuan Lu\thanks{Data Intensive Physical Analytics, IBM Thomas J. Watson Research Center}
\and Youngdeok Hwang\thanks{Department of Statistics, Sungkyunkwan University}}

\maketitle

\vspace{-0.2cm}
\begin{abstract}
Predicting the intensity and amount of sunlight as a function of location and
time is an essential component in identifying promising locations for
economical solar farming.   Although weather models and irradiance data are
relatively abundant, these have yet, to our knowledge,  been hybridized on a
continental scale.  Rather, much of the emphasis in the literature has been on
short-term localized forecasting.  This is probably because the amount of data
involved in a more global analysis is prohibitive with the canonical toolkit,
via the Gaussian process (GP). Here we show how GP surrogate and discrepancy
models can be combined to
tractably and accurately predict solar irradiance on time-aggregated and daily
scales with measurements at thousands of sites across the continental United
States.  Our results establish short term accuracy of
bias-corrected weather-based simulation of irradiance, when realizations are
available in real space-time (e.g., in future days), and provide accurate
surrogates for smoothing in the more common situation where reliable weather
data is not available (e.g., in future years).

\end{abstract}


\section{Introduction}
\label{sec:intro}

Mapping solar irradiance is key to identifying promising locations for solar
power collection. In addition to the obvious economic and energy
sustainability/independence applications, spatial-temporal understanding of
irradiance is also important scientifically.  For example, irradiance is an
important input variable in many biological processes
\citep[e.g.,][]{weiss:hays:2004}. Previous work has so
far been focused on short-term forecasting.   Meteorological variables, such
as temperature, cloud cover, pressure, and wind speed have been successfully used to train
neural networks \citep{alzahrani:dagli:2014,wang:etal:2011} and other
machine learning algorithms \citep{sharma:kakkar:2018} to forecast irradiance and for related tasks. 

However, as far as we know, there has been no published work utilizing
weather-model-based forecasts of irradiance as the basis for such predictions.
Yet such sources abound and, as we show, they are highly
informative.   Examples include simulations furnished by the National Centers
for Environmental Prediction (NCEP) and the European Center for Medium-Range
Weather Forecasts (ECMWF).  Some, like IBM's Physical Analytics Integrated
Data Repository and Services (PAIRS: \url{https://ibmpairs.mybluemix.net}), even
offer simulation-derived output for free through web-based Application Programming Interfaces (APIs). Our goal here
is to wrangle the untapped potential of such sources towards accurate
out-of-sample prediction of observed solar irradiance, i.e., at unknown
locations in space and in space-time.  One challenge in doing so involves
correcting for systematic bias between the simulation model(s) and
unbiased-yet-noisy field observations on a potentially massive scale.

Many scientific phenomena are studied via (computer implementation of)
mathematical models and field experiments simultaneously. Calibrating computer
models to limited field data is a popular topic in the discipline of computer
experiments, usually with abundance in the former compensating for scarcity in
the latter.  Extensive field data collection is often prohibitively expensive
or otherwise impractical, whereas computer simulations are regarded as a cheap
alternative, if not an entirely accurate one owing to idealization or a
limited understanding of the physical dynamics in play.  Computer model {\em
calibration} \citep[e.g.,][]{kennedy:ohagan:2001,higdon:etal:2004} involves
learning the discrepancy between computer models and
field data, via Gaussian Processes (GPs), while ``tuning the knobs'' mapping
simulator output to reality. \citet{kennedy:ohagan:2001}, hereafter KOH,
proposed a now ubiquitous Bayesian framework coupling these two data sources,
which has been applied in a wealth of important practical settings
\citep[e.g.,][]{ling:etal:2012}.

However the KOH framework is imperfectly matched to the synthesis of
irradiance simulations and field observations for three reasons. First, the
data (from either/both sources) is too big for conventional modeling. Second,
the GP is too rigid.  In space, the typical stationarity assumption
can be a poor fit to the underlying dynamics on a continental
scale.  In time the GP is overkill when simpler seasonal
dynamics prevails. Third, no tuning is required in our setup, however
we do need to simultaneously combine multiple simulation sources
\citep{goh:etal:2013}.

Although we retain much of the flavor of the KOH setup, adapting to such
nuances has led us to simplify first: emphasizing modularized
\citep{liu:bayarri:berger:2009} Bayesian calibration via maximization
\citep[e.g.,][]{gramacy:etal:2015} over full sampling-based
ideals. Then we expand to accommodate nonstationarity, temporal dynamics, and
multiple computer models, while attempting to remain computationally
tractable.  The final result is a predictor which leverages computer model
simulations, compensating for systematic discrepancies and/or utilizing
surrogates when model runs are not readily available, that provides an
accurate picture of irradiance at any location in the continental USA in space
and time.  Inverse-variance weighting (IVW) features as a key ingredient in
the combination of forecasts from a multitude of sources: those derived from
the field data alone, and from the two independently calibrated computer
models.  As with other considerations, IVW was conceived as a simple mechanism
in a landscape of more elaborate, but more computationally intensive,
alternatives.

Toward those ends, the remainder of the paper is organized as follows: Section
\ref{sec:dv} describes our solar irradiance data sources in some detail,
providing an overview of the challenges to modeling and synthesis therein.  A
brief review of some of the main components of that methodology, such as GPs
for surrogate and KOH-style discrepancy modeling, is also provided. Section
\ref{sec:ta} describes methods for forecasting with time-aggregated data,
introducing the IVW approach to synthesize a multitude of data sources, and
addressing spatial nonstationarities that arise under the conventional GP
apparatus.  Section \ref{sec:dtie} dis-aggregates over time to explore daily
irradiance in space with a similar setup.  A simple space-time hierarchical
model enables accurate movie-like views into the underlying irradiance
dynamics in space-time. Section \ref{sec:ip} explores the value of additional
simulation runs, via the IBM PAIRS API.   
Finally, Section \ref{sec:discuss} concludes with
brief reflection and perspective.

\section{Data and methodological elements}
\label{sec:dv}

Below we detail data sources, and review modeling elements typically involved
in working with data of this kind.  Many elements of this discussion serve to
motivate enhancement and customization made in our later methodological
sections.

\subsection{Data sources}
\label{sec:data}

\begin{figure}[ht!]
\centering
\vspace{0.3cm}
\includegraphics[scale=0.5,trim=0 125 0 160,clip=TRUE]{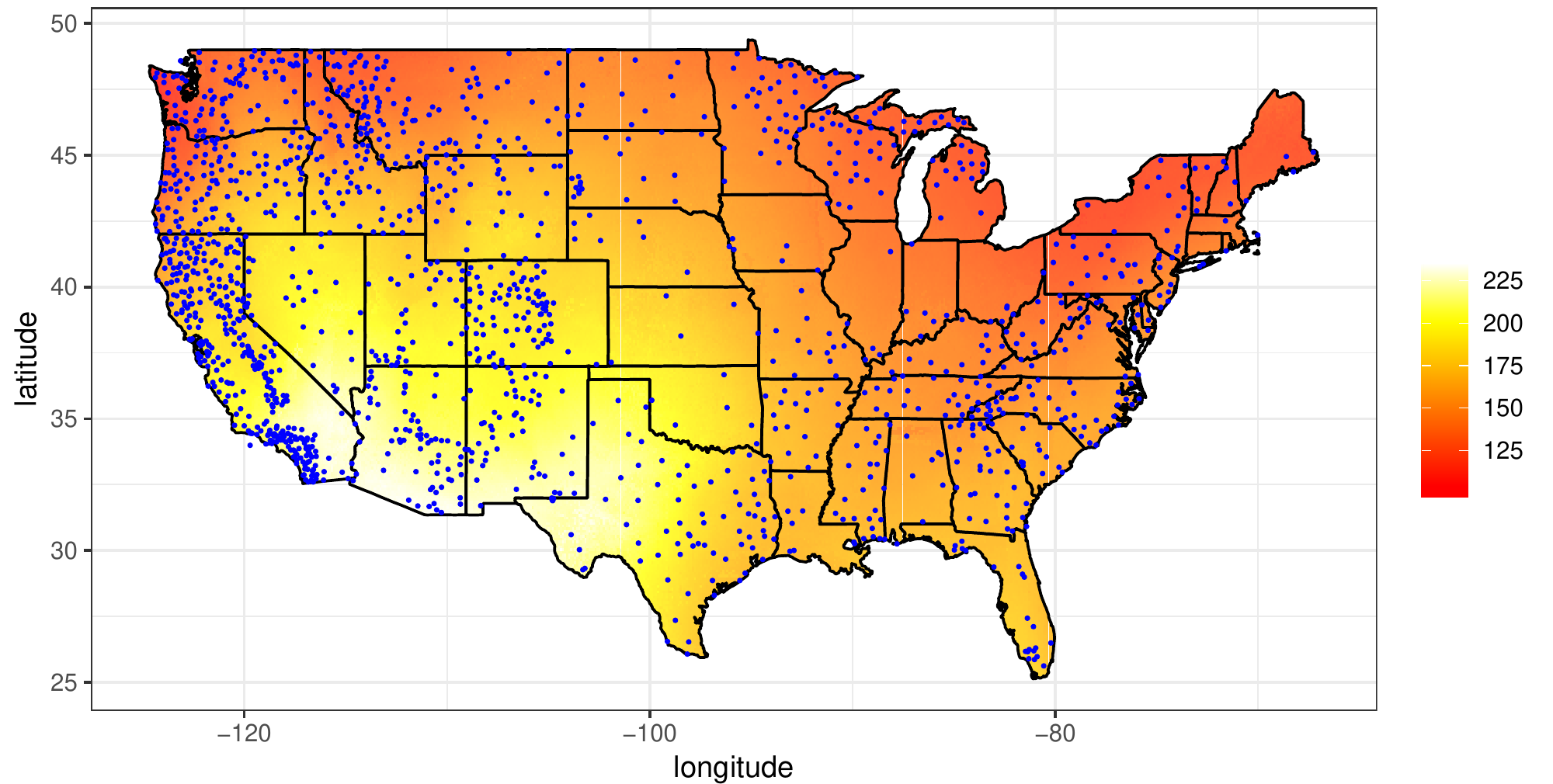}
\caption{Spatial locations of co-located observations of the three data
sources: field, NAM and SREF. A heat map showing prediction of time-aggregated irradiance is overlaid.}
\label{f:dt_orig}
\end{figure}

We have daily solar irradiance measurements from three sources 
co-located at 1535 weather stations distributed throughout the continental United States,
from September 18, 2014 to April 15, 2016. Those stations were taken from selected
sites in the Remote Automated Weather Stations (RAWS) network \citep{zachariassen:etal:2003}.
A visualization of these locations is provided in Figure \ref{f:dt_orig}. 
The forecast of irradiance overlaid on the figure will be discussed in more
detail in Section \ref{sec:ta}.  Observe that the spatial locations of the 
RAWS sites are not evenly distributed. The network is run by the U.S. Forest Service and 
Bureau of Land Management and thus sites favor locations in national forests.  
The daily resolution of the data over this time period means that we have
almost 900 thousand observations from each data source.  The measurements we
use are integrated from sub-hourly or hourly samples, but we do not work directly
with data on that resolution for a variety of reasons, the most important
being size.  Another is a substantial degree of missingness:  about 17\% of
locations are missing more than 5\% of daily observations, and no location is
fully observed. Many RAWS sites are located in remote areas---often 
not connected to the electric grid and relying on satellite links for communication---and thus
failure/outages are common.

The field data, which are measurements of global horizontal irradiance (GHI) 
provided by equipment housed at those weather
stations, was retrieved from the MesoWest project run by 
the Department of Atmospheric Sciences at the University of Utah 
\citep[\url{mesowest.utah.edu}][]{Horel:etal:2002}. Our simulation data
come from two sources which leverage numerical weather models that
are conditioned on historical observed weather patterns.  So whereas the field
data are sensitive to observed weather, our simulations depend on weather
forecasts.  Clearly the former will be noisier than the latter.
Although the latter are technically deterministic, given forecasted weather,
there is inherent uncertainty in those forecasts which is important to
acknowledge when thinking about a typical day (for which weather forecasts are
not readily available) compared to a historical one (for which there are).

The first simulation-based irradiance source comes from the North 
American Mesoscale (NAM) model using a 5 km Lambert conformal grid, a public resource provided 
by the NCEP \citep{rogers:etal:2009}, 
and made available by the National Oceanic and Atmospheric Administration (NOAA).  
NAM irradiance forecasts are derived from observed and forecasted meteorological 
variables such as temperature, precipitation, etc. Our sampling of the NAM data comes from IBM PAIRS which
provides a convenient API for an abundance of simulation data, as well as satellite imagery,
soil and land data and sensor measurements, curated into an integrated form,
projected onto a common coordinate system, and indexed for convenient
probing/downloading \citep{klein:etal:2015,lu:etal:2016}. Section \ref{sec:ip} 
provides greater detail on our use of PAIRS, in particular to augment our 
corpus of simulation data in order to show the benefit of relatively easy
access to this valuable data source. 

Our second simulation derives from the Short Range Ensemble Forecast (SREF)
advanced research Weather Research and Forecasting (WRF) core, with a 40 km Lambert Conformal grid \citep{lu:etal:2016}. 
SREF uses an ensemble of perturbed initial conditions into the underlying 
numerical atmospheric models \citep{taylor:buizza:2002}, which include NAM-derived simulations.
It has been demonstrated that this ensemble approach better captures observed variability 
in forecasts so-obtained, compared to simpler ``point-wise” forecasts 
\citep{krishnamurti:etal:2000}. 

Our goal is to synthesize these three data sources in order to accurately
capture the evolution of solar irradiance over time and space, focusing on the
continental USA, and in particular on the regions which are under-sampled in
space.  Below, we review Gaussian process (GP) models, which are canonical in
both spatial smoothing and computer surrogate modeling contexts, and
\citeauthor{kennedy:ohagan:2001} (KOH) style discrepancy modeling for bias
correction.  Addressing challenges in utilizing such machinery in our data
context, mostly to do with data size and requisite modeling fidelity, comprise
of the main motivations for our methodological work.

\subsection{Gaussian process review}
\label{sec:gp}

Gaussian processes (GPs) have a rich tradition in spatial modeling and
geostatistics, going back to \citet{math:1963} and with excellent technical
\citep[e.g.][]{stein:1999} and methodological \citep[]{cressie:1993}
resources.  More recently, they have played a major role as strong
nonparametric predictors in the design and analysis of computer experiments
\citep{sacks:etal:1989,sant:will:notz:2003} where they serve as surrogate
models or emulators replacing expensive numerical calculation for the modeling
of (mostly physical) phenomena; and in the machine learning literature
\citep{rasmu:will:2006} where they tend to dominate in out-of-sample
prediction exercises in settings exhibiting smooth and high-signal input-output
relationships.  Technically speaking, a GP is a prior over random functions,
where any finite collection of function evaluations
follow a multivariate normal (MVN) distribution.  Thus those realizations are
fully described by the mean function $\mu(x)$ and positive-definite covariance
function $\Sigma(x,x')$ for any pair of spatial locations $x$ and $x'$. With
data $D_N = (X_N, Y_N)$,
where $X_N$ is an $N\times p$ matrix of inputs and $Y_N$ is an
$N$-sized response vector, the GP prior implies an MVN sampling model, $Y_N
\sim \mN_N(\mu_N,\Sigma_N)$ with $\mu_N$ and $\Sigma_N$ are defined
by applying the mean and variance functions to elements of $X_N$.

Whether the problem is spatial, with $x$ being geo-located coordinates, or
applied more generally in surrogate modeling or machine learning contexts, the
emphasis is on regression: deriving the conditional distribution of $Y(x) |
D_N$, which follows immediately from MVN conditioning identities.  If $\mu_N =
0$, a common simplifying assumption, and $\Sigma(X_N,x)$ is the $N \times 1$
matrix comprised of $\Sigma(x_1, x),
\dots, \Sigma(x_N, x)$, then $Y(x) \mid D_N \sim
\mathcal{N}(\mu(x), \sigma^2(x))$, with
\begin{align}
\mbox{mean } \quad \mu(x) &= \Sigma(x, X_N) \Sigma_N^{-1} Y_N \nonumber \\
\mbox{and variance } \quad \sigma^2(x) &= \Sigma(x,x) - \Sigma(X_N, x)^\top \Sigma_N^{-1} \Sigma(X_N, x). \label{eq:gppred}
\end{align}
It is clear from the equations above that the covariance function $\Sigma(x,
x')$ plays a crucial role, and that computational complexity is cubically
linked to $N$, which could severely limit training data sizes in practice.
Common choices of $\Sigma(x, x')$ encode prior beliefs about the function
spaces spanned by the predictive equations, particularly their smoothness and
decay of correlation as a function of input distances.   They are often
parameterized, and determining settings for those parameters is additionally
fraught with computational challenges, requiring repeated cubic decomposition
of $N \times N$ covariance matrices $\Sigma_N$. Details on the libraries used for
inference, and thereby the precise correlation structure, etc., are provided
later alongside our empirical work.

In addition to being computationally cumbersome in large-$N$ contexts, which
will be problematic with $N$ in the thousands (or hundreds of thousands), the
typical choices for $\Sigma$ are overly rigid in their assumption of
stationarity, or more specifically of translation invariance in space:
$\Sigma(x, x') = \Sigma(x - x')$, being a function only of the input
displacement $x - x'$.  It would be quite surprising to find that spatial
dependence (i.e., spatial correlation) in irradiance on the West of the US is
the same as in the Midwestern Plains, or on the East Coast.  Although there
are many approaches to non-stationary GP modeling in the literature
\citep[e.g.][]{paciorek:schervish:2006,schmidt:ohagan:2003}, few manage
without dramatically expanding the requisite computational demands.  Methods
which divide-and-conquer \citep[e.g.][]{kim:mall:holm:2002,gra:lee:2008},
simultaneously leveraging statistical (and thus geographical) and
computational independence, are an important exception.  In Section
\ref{sec:ta} we explore a particularly attractive recent divide-and-conquer
method in this context.  Space-time modeling faces similar large data and
modeling fidelity challenges, especially in the face of missing data, which we
address with a remarkably simple apparatus in Section \ref{sec:dtie}.

\subsection{Learning discrepancies}
\label{sec:discrep}

Modeling the bias in our computer simulations, i.e., the discrepancy between
field data observations and simulations, will be a recurring theme in the
development of (our best) irradiance predictors.  We do this by borrowing the
calibration apparatus of KOH, but without the need to calibrate an unknown
tuning parameter.  We model field observations as connected to
computer model output via
\begin{equation}
Y^F(x) = Y^M(x) + b(x) + \epsilon, \quad \text{where} \quad 
\epsilon\overset{\mathrm{iid}}{\sim} \mathcal N(0, \sigma^2_\epsilon).\label{eq:fc}
\end{equation}
Above, $F$ denotes the field process, $M$ a (computer) model, and $b$ the
bias correction. 

As described in more detail in later sections, we entertain variations which
utilize computer model realizations $Y^M(x)$ directly, i.e., in the style of
\citet{higdon2004combining}, and ones which lean on a surrogate model
$\hat{y}^M(\cdot)$ fit to simulator data.  In this latter case, we depart from
KOH and modularize \citep{liu:bayarri:berger:2009} so that $\hat{y}^M(\cdot)$
is fit independently of $\hat{b}(\cdot)$, which is sensible (and
computationally advantageous) in situations where the amount of field and
simulation data is comparable. We keep within KOH framework with GP-based
priors on those quantities, however we avoid computationally intensive MCMC to
prefer derivative-based posterior maximization via residuals $Y_N^F -
\hat{Y}^M_N$ to train $\hat{b}(\cdot)$, following \citet{gramacy:etal:2015}.
%
Our setup applies this apparatus separately for each computer model.  Bias corrected
predictions are then combined via the cascade of GP predictive variances
(\ref{eq:gppred}) involved in each fit.

\section{Time-aggregated exploration}
\label{sec:ta}

We begin by describing the simplest out-of-sample (OOS) prediction exercise
we could imagine in the context of predicting solar irradiance with the data
that we have. Then throughout this section and the next two we gradually
increase the fidelity of the model and resolution of the data in pursuit of
higher accuracy and more precise forecasts.  In particular, this section
focuses on time-averaged data. For each of 1535 spatial locations, we
work with average irradiance values from both the field, and the NAM and SREF
computer models.  

In that context, consider the following leave-one-out cross validation (LOO-CV
for short) exercise:  iteratively hold out a particular spatial location,
train on the data remaining at the other 1534 locations, and make a prediction
for the held-out value.  There is some variation in how computer model and
field data are held out in tandem which will be further detailed below.
We measure accuracy of each prediction in terms of root-mean-squared error
(RMSE), collecting 1535 values in total.  Although RMSE in this case is the
same as mean absolute error, since only one (time aggregated) value is held
out, we refer to RMSE here in order to better connect with time dis-aggregated
results in later sections. The full suite of results from this exercise is shown
in Table \ref{t:loo_cv}, where the columns and rows of the table traverse the
cascade of methods described in more detail below.

\begin{table}[ht!]
\centering \small
\begin{tabular}{rr|r|rrrrrr|r}
& target & data/model & global & 95cov & $p_{\mathrm{global}}$ & local & 95cov & $p_{\mathrm{local}}$ & $p_{\mathrm{locvglob}}$\\ 
  \hline
1& field& $\widehat{\mathrm{field}}$& 34.88& 0.9485& & 24.73& 0.9440 && 0\\ 
2& NAM&  $\widehat{\mathrm{NAM}}$ & 23.22& 0.9629& & 9.69& 0.9446 && 0\\ 
3& SREF&  $\widehat{\mathrm{SREF}}$ &22.27& 0.9603 & & 9.99& 0.9479 && 0 \\ 
\hline 
4& field&  $\widehat{\mathrm{SREF}}$ no b& 45.63& 0.1857& &45.43& 0.1752& &0.6605\\ 
5& field&  $\widehat{\mathrm{NAM}}$ no b& 33.62& 0.4436& 0.9004$^1$& 33.48& 0.3511& &0.3857\\ 
6& field&  $\widehat{\mathrm{NAM}}$ + $\widehat{\mathrm{b}}$& 25.30& 0.9629& 0$^5$& 24.70& 0.9485& 0.2305$^1$& 0.01624\\ 
7& field&  $\widehat{\mathrm{IVW}}$& 25.09& 0.8391& 0.05625$^6$& 24.65& 0.8156& 0.4069$^6$& 0.09994\\ 
\hline
8& field &  SREF no b&  44.97 & & & & &&\\
9& field &  NAM no b& 32.80 & & & & &&\\ 
10& field&  NAM + b& 24.07& 0.9577& 0.01666$^7$& 23.68& 0.9472& 0.06869$^7$& 0.3811\\ 
11& field&  IVW& 23.63& 0.8326& 0.5444$^{10}$& 23.41& 0.8104& 0.03503$^{10}$& 0.006098\\ 
\end{tabular}
\caption{LOO-CV average RMSE and 95\% out-of-sample predictive coverage.  The $p_{\mathrm{global}}$ and
$p_{\mathrm{local}}$ columns summarize the output of a one-tail paired
$t$-test (to be detailed in Section \ref{sec:global}) to the next best method (for the same target) lying in 
the previous column but a higher row in the table.  Log values are used.  The superscript
indicates the row being compared against.  The $p_{\mathrm{locvglob}}$ column
shows a similar result for the best versus the second-best (i.e., local
v.~global) in the same row of the table.  A zero $p$-value in any column is a
shorthand for $<2e^{-16}$ output from the software.} 
\label{t:loo_cv}
\end{table}

\subsection{Global models}
\label{sec:global}

We first consider forecasts based on ordinary GP models, as reviewed in
Section \ref{sec:gp}, which we refer to here as ``global'' for reasons which
will be revealed in the next subsection, entertaining local models.  To train
the GPs we considered an implementation in the {\tt mlegp} library
\citep{mlegp} on CRAN \cite{cranR}, and one via the {\tt mleGPsep} function in
the {\tt laGP} library \citep{gramacy:2014,gramacy:jss:2016}, also on CRAN,
additionally implementing the local add-ons below.  Both gave remarkably
similar results despite differences in how the mean function is parameterized.
These RMSEs reside in the ``global'' column of Table \ref{t:loo_cv}.

Start with the first row (and ``global'' column) in Table \ref{t:loo_cv}, which
compares predictions derived solely from the field data.  That is, a GP was
trained on 1534 input-output field data pairs, leading to
$\widehat{\mathrm{field}}$, and used to predict the held-out observation.  The
target column, indicating ``field'' in this case, somewhat redundantly
clarifies that we are predicting the held-out field data observation.  Other
variations will be entertained shortly.  After looping over all 1535 
folds we obtain an average RMSE of 34.88. This number will serve as the
baseline for further experimentation. As a proportion of the total range
of aggregate irradiance observations, which is zero (an erroneous
value for a daily aggregate) to about 317, that RMSE represents about 11\%
of the range.

We wish to improve upon this number by incorporating simulations from NAM
and SREF.  Toward that end we trained 1535 LOO-CV GP predictors, of exactly the
same sort used in our ``global'' $\widehat{\mathrm{field}}$ fits in the
paragraph above, separately to the NAM and SREF observations.  The next two
rows in the table show the accuracy of these predictors, labeled
$\widehat{\mathrm{NAM}}$ and $\widehat{\mathrm{SREF}}$ respectively.  Clearly
these predictors are of higher quality than the $\widehat{\mathrm{field}}$
analogue.  This makes sense on an intuitive level: the computer model
simulations are less noisy (involve more signal) than the field data
observations, so they are easier to predict with the same library.  However,
that comparison is not of direct interest. The NAM \& SREF simulators, and
correspondingly their trained surrogates, are only useful insofar as they
enable us to better predict the actual (field) irradiance at a particular
location.

Rows 4--5 show how accurate the NAM and SREF emulators are directly.  That is,
we treat the $\widehat{\mathrm{NAM}}$ and $\widehat{\mathrm{SREF}}$ forecasts
as ``field'' forecasts, and they are not particularly good: NAM offers slight
improvement, SREF is actually worse, compared to our earlier
$\widehat{\mathrm{field}}$ results. The $p_{\mathrm{global}}$ value
immediately to the right in the table demonstrates $\widehat{\mathrm{NAM}}$'s
improvement over $\widehat{\mathrm{field}}$ predictions is not statistically
significant. Throughout, pairwise $t$-tests compare differences in log squared
error across holdouts for pairs of models. The two sets of errors in question
reside on the row of the quoted $p$-value, and the commensurate row above is
indicated by a superscript.  If $A$ is the former, $B$ is the latter, and
$\mu_{(\cdot)}$ are their average log squared errors, then the hypotheses
involved are $\mathcal{H}_0: \mu_A\geq \mu_B$, versus $\mathcal{H}_1:
\mu_A < \mu_B$. Perhaps the reason for ``$\widehat{\mathrm{NAM}}$ no b" not being better than
$\widehat{\mathrm{field}}$ is that the computer models are biased (``no b''
means ``without bias correction'').  To investigate, we separately fit two
discrepancy terms, as described in Section \ref{sec:discrep}. Specifically, we
fit a GP to the residual between in-sample surrogate predictions, from
$\widehat{\mathrm{NAM}}$ and $\widehat{\mathrm{SREF}}$ respectively, leading
to $\hat{b}_\mathrm{NAM}$ and $\hat{b}_\mathrm{SREF}$.  Row 6 in the table
shows average RMSE for the resulting bias-corrected ``$\widehat{\mathrm{NAM}}$
+ $\hat{b}_\mathrm{NAM}$'' comparator, with the SREF analogue being similar
and thus omitted.  Observe that this is much improved compared to the
un-bias-corrected version, and that the $p_{\mathrm{global}}$ column reveals
that this result is highly statistically significant.

Now before discussing the final row in that block (row 7), which combines the
NAM and SREF (bias-corrected) emulators, it is worth discussing the subtle
variation reported in the final block.  These forgo emulation of the computer
models and condition on the true values output from the simulator, presuming
ad hoc access to the simulator were readily available (including all of the
relevant weather data required as input to those models).  Although this is
somewhat unrealistic, such a comparison is useful as a gold standard.  The
first two rows in that section (rows 8--9 from the top), which do not correct
for bias, perhaps suggest that such simulations are not directly useful.
However the final lines in the table/block suggest that bias correction here
yields the most accurate predictions obtained so far, if by a relatively small
margin.  Note that although there is no hat on the ``b'' in these entries of
the table, the bias is still estimated by training GPs on residuals in the
same way. 

\subsection{Inverse-variance weighting}
\label{sec:ta_ivw}

It is natural to ask how all three data sources, and
their bias corrections, could be combined to potentially emit a predictor
which is more robust than any of the three alone.  This has a spirit similar to 
multi-fidelity
calibration \citep[e.g.][]{goh:etal:2013}.  However that is an imperfect
analogy as the notion of fidelity is weak and potentially misleading. Although
the SREF ensemble is ``bigger than NAM'', it isn't exactly a finer resolution
alternative. Rather, it is perhaps a more conservative (and thus a potentially lower
resolution) alternative despite being based on a more aggressive simulation
effort.

Rather than attempt to jointly model all unknowns, as is common in the
multi-fidelity calibration literature, we take the simpler tack combining
``independent'' estimators.  We use the term ``independent'' loosely in this case,
since all three data sources are derived from similar measurements, however
they are gathered/calculated from physically independent sources.  In the
context of combining statistically independent (and unbiased) estimators, an
additive inverse-variance weighting (IVW) scheme is optimal, minimizing 
the variance of the weighted average \citep{cochran:1954}.  That is, we
consider
\[
\hat{y}(x) \propto \sum_j w^j(x) \hat{y}^j(x)\mbox{, where in our particular
case} \ j \in
\{\widehat{\mathrm{field}}, \widehat{\mathrm{NAM}}, \widehat{\mathrm{SREF}}\},
\]
and the normalizing denominator comes from summing all weights for each
particular $x$. Each $\hat{y}^j(x)$ is derived from the same
calculations used in other rows of Table \ref{t:loo_cv}, and the $w^j(x)$ are
proportional to the inverse variance calculated for the respective prediction.
In particular, $\hat{y}_{\widehat{\mathrm{field}}}(x)$ is simply a calculation of
$\mu_{\widehat{\mathrm{field}}}(x)$ following Eq.~(\ref{eq:gppred}), and the
corresponding weight is $w^{\mathrm{field}}(x) =
1/\sigma_{\widehat{\mathrm{field}}}^2(x)$.  The predictions and weights from
the bias adjusted variations require summing two sets of means and
(inverse) variances from the surrogate and bias correction GPs, respectively.
In the case of the unrealistic comparators at the bottom of the table, which
condition on known computer model simulations, the mean and weight are
derived from the bias-corrected GP only. 

Observe from Table \ref{t:loo_cv}, specifically rows 7 and 11, that these IVW
comparators offer slight improvements over analogues earlier in the table.  It 
would appear that most of the heavy lifting is being done by the 
``$\widehat{\mathrm{NAM}} + \widehat{\mathrm{b}}$'' mixture component.  However, 
we draw comfort from a fuller incorporation of uncertainty, especially as regards the 
incorporation of the more conservative SREF predictions.
Figure \ref{f:dt_orig} shows the resulting $\hat{y}(x)$
values of this IVW surrogate-based version (row 7 from the table) overlaid
onto a map of the USA, with the training locations shown.  This
predictor was derived from a single application on all 1535 data locations,
not from the LOO-CV exercise summarized in the table.  The three sets of
predictive equations, and their corresponding weights, were evaluated at a
dense grid of about eighty thousand locations spread evenly throughout the
continent.  The alternative in row 11 of the table is not an option
here, as we do not have the luxury of obtaining NAM and SREF on such a dense
grid.  In Section \ref{sec:ip} we consider some  adaptations that might be
possible if new runs were more readily available.

\subsection{Coping with continental nonstationarity}
\label{sec:laGP}

The preceding analysis was based on stationary GP models, which are the default
in the literature. These models involve a decay of spatial correlation that is
constrained to be identical throughout the input domain.  A
myriad of geological features, such as large mountain ranges, more gradual
changes in elevation (which are not recorded in our data), lakes, forestation,
etc., suggest heterogeneity rather than homogeneity in the input-output relationships
over large spatial scales.  It is also likely that the effects of such
features vary with latitude---with the angle of the sun being a key factor
in solar intensity.

Here we entertain relaxing stationarity as a prelude to a more nuanced spatial
(and spatio-temporal) analysis provided in the following section.  Although
many methods of relaxing stationarity have been suggested in the literature,
from warping stationary models
\cite[e.g.][]{paciorek:schervish:2006,schmidt:ohagan:2003} to (Bayesian)
domain partitioning \citep[e.g.,][]{kim:mall:holm:2002,gra:lee:2008}, few
offer software {\em and} computational tractability as a feature in the face
of large $N$.  An exception is the recently developed local approximate 
Gaussian process developed by \citet{gramacy:apley:2015} in the {\tt laGP} 
package \citep{gramacy:2014,gramacy:jss:2016} on CRAN.

The basic idea of {\tt laGP} involves an implicit introduction of sparsity into
the covariance structure through a {\em transductive} learning technique that
tailors approximate predictive equations to the input locations, $x$, at which
they are desired.  Although each $x$ entertains the full data, $D_N =
(X_N, Y_N)$, it only does so indirectly via a search for sub-design $X_n(x)$
primarily comprised of $X_N$ ``close'' to $x$.
\citeauthor{gramacy:apley:2015} show how several variations on a greedy search
for these local designs yield local predictors based on $D_n(x) = (X_n(x),
Y_n(x))$, with $n \ll N$, which are at least as accurate as simpler
alternatives (such as nearest neighbor) yet require no extra computational
time.  They continue to show how the method can be applied to a vast set of $x\in
\mathcal{X}$, massively parallelized when such architectures are available
\citep{gramacy:niemi:weiss:2014}, and how the result takes on a nonstationary
flavor where spatial dependence can vary throughout the input domain.  Hybrid
versions developed by \citet{sun:etal:2017}, which essentially combine
ordinary ``global'' GPs applied to (computationally manageable) subsets of the
data with (full data) {\tt laGPs} on the residuals, have led to competitive
results compared to other publicly available software \citep{heaton:etal:2017}.

We entertained {\tt laGP} variations for predicting solar irradiance in our
LOO-CV setup, simultaneously as surrogates and models for discrepancy, and the
results are summarized in the final columns of Table \ref{t:loo_cv}. Those
toward the top of the table, showing accuracy on simple prediction exercises
with $\widehat{\mathrm{field}}$, $\widehat{\mathrm{NAM}}$ and
$\widehat{\mathrm{SREF}}$, offer the most stark contrasts.  Here {\tt laGP}
provides reductions in RMSE of between thirty and fifty percent.  Those
increases in accuracy translate to the bias-corrected variations lower in the
table, but the gap in prowess is somewhat ``washed out'' as more (hitherto
stationary) GP models are combined. Observe that those IVW-based global
models generally have better out-of-sample coverage at the nominal 95\%
level. The final column summarizes pairwise $t$-tests across the rows of the
table, pitting the comparator with the best average RMSE against the second
best one in that row.  The take-home message is that the local methods offer a
small improvement over the global ones, but one that is statistically
significant for the best methods being entertained: IVW combined schemes,
either via actual simulations or in the more realistic setting where
surrogates are used.  A referee suggested that we add a Bayesian
additive regression tree \cite[BART,][]{chipman:2010} comparator, which is
provided in Appendix \ref{sec:bart}.

\section{Daily time-incorporated output}
\label{sec:dtie}

\begin{figure}[ht!]
\centering
\includegraphics[scale=0.35,trim=0 10 30 25,clip=TRUE]{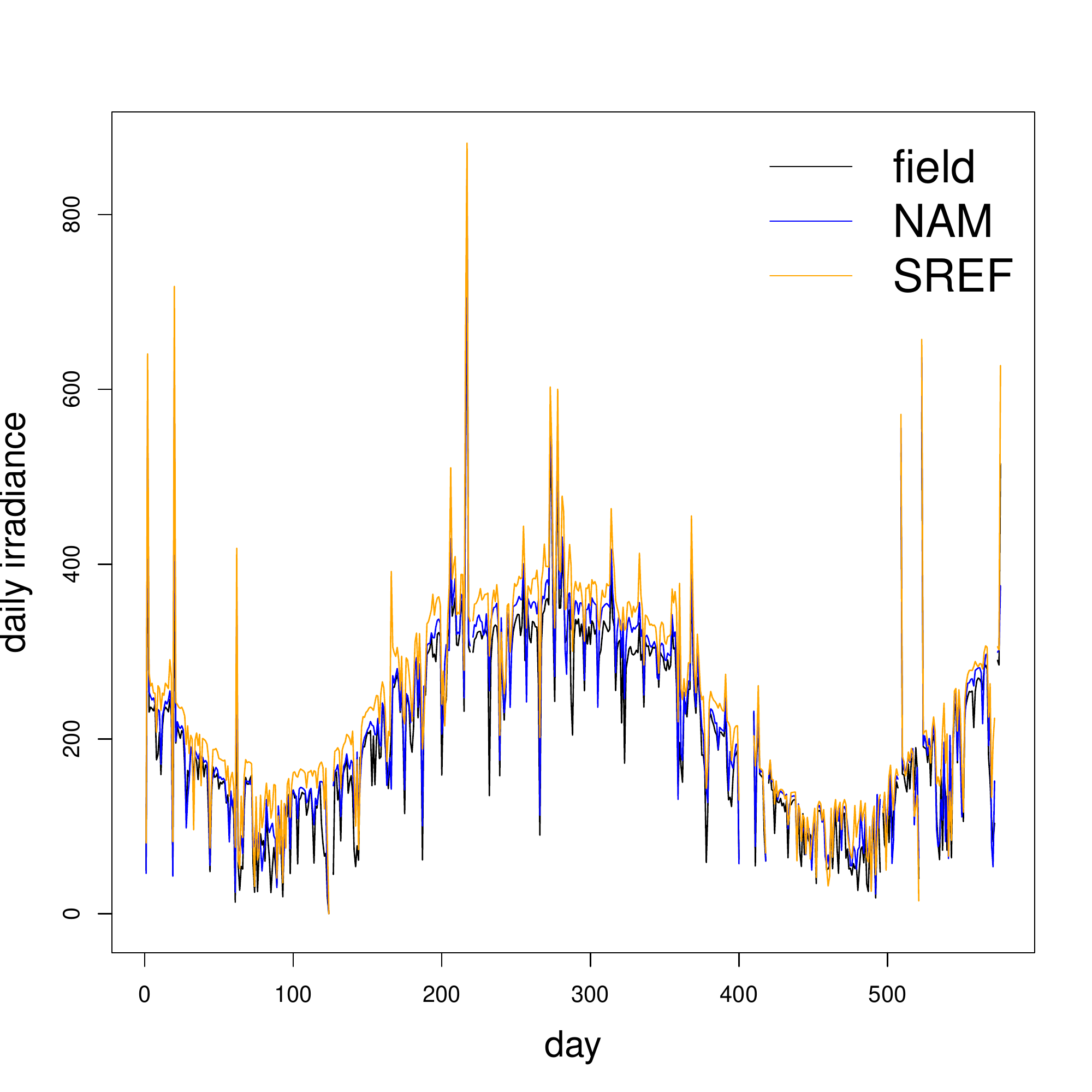} \hspace{0.5cm}
\includegraphics[scale=0.35,trim=30 10 30 25,clip=TRUE]{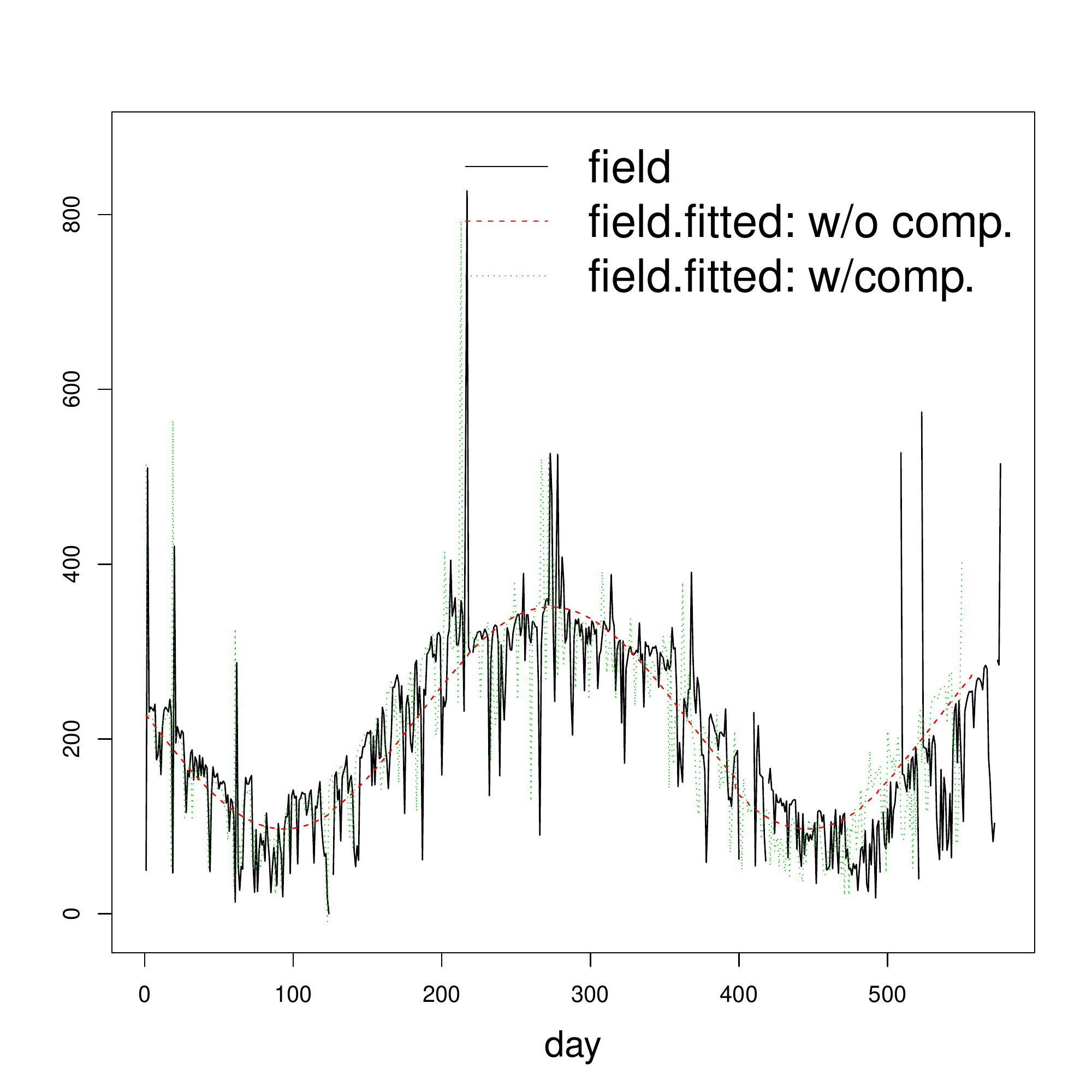}
\caption{{\em Left}: daily solar irradiance field measurements, NAM and SREF
         simulations, respectively, at (lat, lon) $= (37.69, -121.6)$; {\em Right}:
         output of a simple linear (autoregressive and periodic) time series
         regression fit.}
\label{f:dtie}
\end{figure}

We now move to a space-time analysis on the (dis-aggregated) daily resolution
of data.  To get a feel for how the data look in time, the left panel of
Figure \ref{f:dtie} shows our daily observations of the field data and two
computer models for (lat, lon) $=(37.69, -121.6)$ in Northern California.
It is clear from this view that there is substantial correlation between these
observations, and a clear (predominantly periodic) pattern in time. 

One way to learn relationships between such covariates, namely between time
and the set of three field and simulation measurements (i.e., ignoring the
spatial component for the moment), is through linear (time series) modeling.
For example, consider the following model
\begin{equation}
Y^F(t) = \beta_0 + \beta_1 \sin\left(\!\frac{2\pi t}{365}\!\right) + \beta_2 \cos\left(\!\frac{2\pi t}{365}\!\right) + 
\beta_3 Y^{\mathrm{F}}(t-1) + \beta_4 Y^{\mathrm{NAM}}(t) + \beta_5 Y^{\mathrm{SREF}}(t) + \varepsilon_t
\label{eq:tsm}
\end{equation} 
with $\varepsilon_t \stackrel{\mathrm{iid}}{\sim} \mathcal{N}(0, \sigma^2)$.
It would be challenging to fit such a model for all spatial locations owing to
the degree of missingness (which was easy to ignore in the time-aggregated
analysis of Section \ref{sec:ta}).  Location (lat, lon) $=(37.69, -121.6)$ is
almost fully observed in the field, missing seventeen of 559 days of
measurements. 
The right panel of Figure \ref{f:dtie} shows the fitted values
of such a regression overlaid by the true measurements.
The model was fit via the {\tt lm} command in {\sf R},
and all estimated coefficients $\hat{\beta} = (\hat{\beta}_0,
\hat{\beta}_1, \dots, \hat{\beta}_5)$ are statistically significant.  
The fit is highly accurate: periodic and
autoregressive predictors, combined with computer model simulation output,
lead to highly accurate ``forecasts'' of observed irradiance.

We put ``forecasts'' in quotes because it doesn't represent a realistic
scenario.  To apply such a model in practice, e.g., forecasting forward 
in time, would additionally require (i.e., on top of a completely observed
historical data set) computer simulation forecasts forward in time (which 
we do not have) and a propagation of uncertainty due to the autoregressive
component.  Incorporating spatial correlation would involve a substantial
degree of added complexity.  We therefore punt on the idea of obtaining 
such forecasts, forward in time, and concentrate instead on extending the 
style of analysis provided by our time-aggregated data in Section \ref{sec:ta}: to
smooth, summarize, extrapolate and correct for biases in predictions in
space-time for a ``typical year''.  The details are provided below.

\subsection{Spatial regularization of seasonal smoothers}
\label{sec:tss}

Let $s\in \mathcal{S}$ index the two-dimensional spatial coordinates of our
observations (with $|\mathcal{S}| = N = 1535$), $t$ index time, and as
previously let $j \in \{\mathrm{field}, \mathrm{NAM}, \mathrm{SREF}\}$ select
the data source.  Now consider the following hierarchical model offering a
spatial regularization of local periodic smoothers.  Independently for each
data source $j$, take
\begin{align}
Y^j_{st} &\sim \mathcal{N}\left[\beta_{0s}^j + \beta_{1s}^j \sin\left(\!\frac{2\pi t}{365}\!\right) \label{eq:hier}
+ \beta_{2s}^j \cos\left(\!\frac{2\pi t}{365}\!\right), (\sigma_s^j)^2  \right] & \forall t \; &\mathrm{s.t.} \; y^j_{st} \;\; \mbox{is recorded} \\
\beta_{ks}^j &\sim \mathcal{GP}^j(s) & k&=0,1,2. \nonumber
\end{align} 
In other words, we propose modeling the time dynamics at each
spatial location $s$ with a simple intercept-adjusted yearly-seasonal linear
model.  However, instead of treating each spatial location independently of
the next we deploy a GP prior on the coefficients $\beta_s^j = (\beta_{0s}^j,
\beta_{1s}^j, \beta_{2s}^j)$ to encourage these to evolve smoothly in
space, and thus borrow information from locations nearby.  By
$\mathcal{GP}^j(s)$ we simply mean to apply a GP (separately for each data
source $j$) with inputs $X_N$ comprised of the corpus $s\in \mathcal{S}$,
exactly as in Section \ref{sec:ta}.

In fact, observe that the GP model(s) in Section \ref{sec:ta} are a special
case of (\ref{eq:hier}) where the periodic coefficients are dropped:
$\beta_{1s}^j =
\beta_{2s}^j = 0$ for all $s$, allowing only the intercept $\beta_{0s}^j$,
representing the average level of the data collected in time, to vary
spatially.  It is worth noting that Eq.~(\ref{eq:hier}) is not a fully
Bayesian hierarchical model, but it wouldn't be hard to complete such a
specification with appropriate priors on the variances $(\sigma_s^j)^2$ and
any hyper-parameter priors on the GPs.  We don't bother here because we don't
believe that inference in such a framework is computationally tractable,
whether fully Bayesian or otherwise, say via posterior maximization. In that
context the $\beta_{s}^j$ vectors are high dimensional latent variables,
having on the order of $3 \times 1535 = 4605$ settings that each would require
integrating over via MCMC (or maximizing over via numerical methods), with
each iteration requiring likelihood (and possibly derivative) evaluations that
are cubic in $N = 1535$.

Instead we take the following vastly simplified, and far more computationally
manageable, approach to make  inference.  We first perform separate Ordinary Least Squares (OLS) regressions,
corresponding to the top line in Eq.~(\ref{eq:hier}), for each spatial
location $s$.  The result is a collection of three-vectors
$\{\hat{\beta}_s^j\}_{s\in \mathcal{S}}$. Then we train three GPs, one for
each coordinate of $\hat{\beta}_s^j$, where the $y$-values in the training set
are those very same $\{\hat{\beta}_{si}^j\}_{s\in \mathcal{S}}$ values stacked
as an $|\mathcal{S}|=N$-vector, separately for each $i \in \{0,1,2\}$, and the
inputs $x$ are the commensurately stacked spatial locations $s \in
\mathcal{S}$.  Next, the predictive equations (\ref{eq:gppred}) provide
forecasts $\tilde{\beta}(\tilde{s})$ at new locations $\tilde{s}$, and furnish
fitted values defining the residuals used to train discrepancies, discussed
momentarily.  Forming predictions for irradiance outputs in time involves
pushing these smoothed $\tilde{\beta}_s^j$ back through the first line in the
hierarchical model (\ref{eq:hier}) for the $t$-values (i.e., days) of
interest.

\begin{figure}[ht!]
\centering
\hfill \includegraphics[scale=0.25,trim=30 30 0 0]{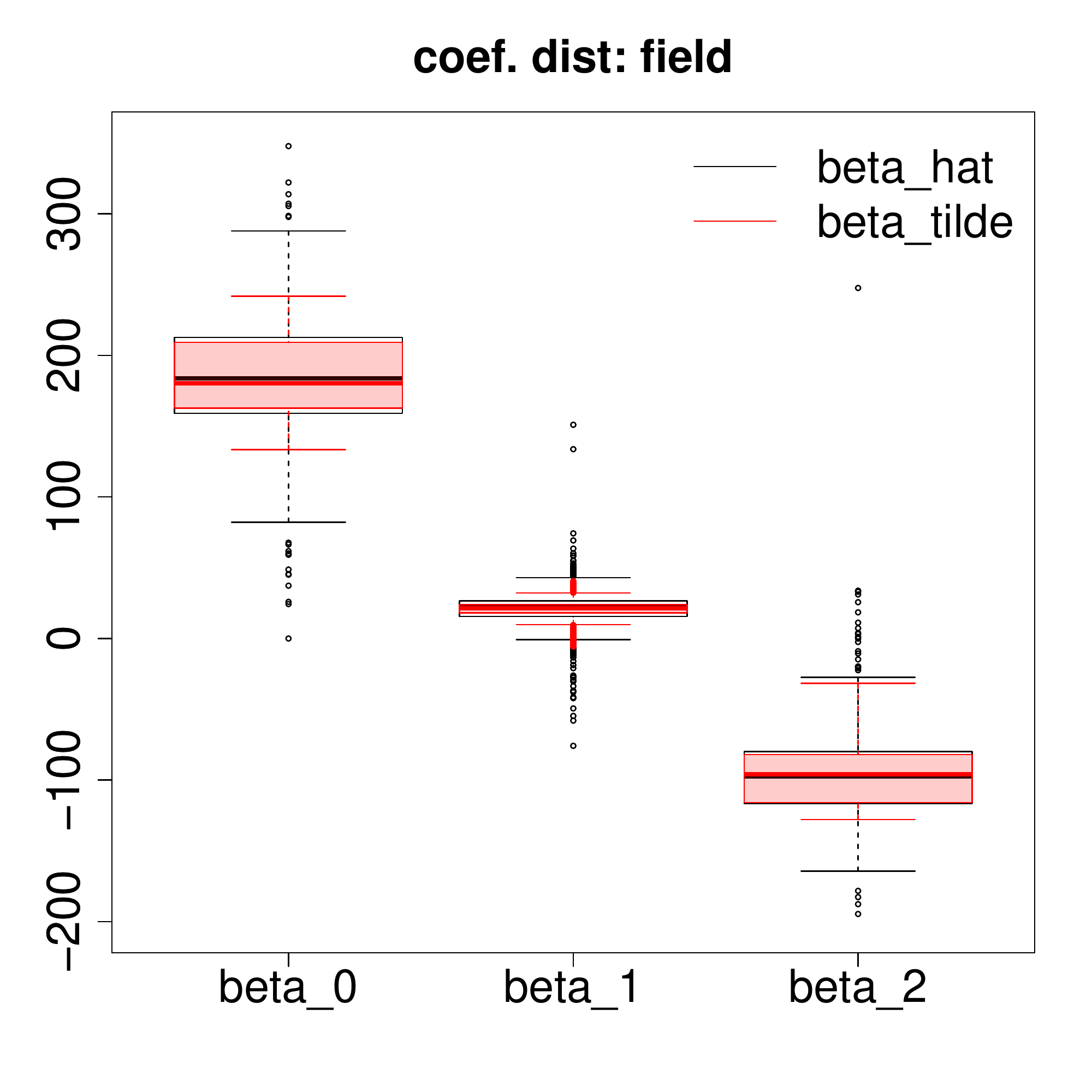} \hfill
\includegraphics[scale=0.4,trim=0 120 0 120,clip=TRUE]{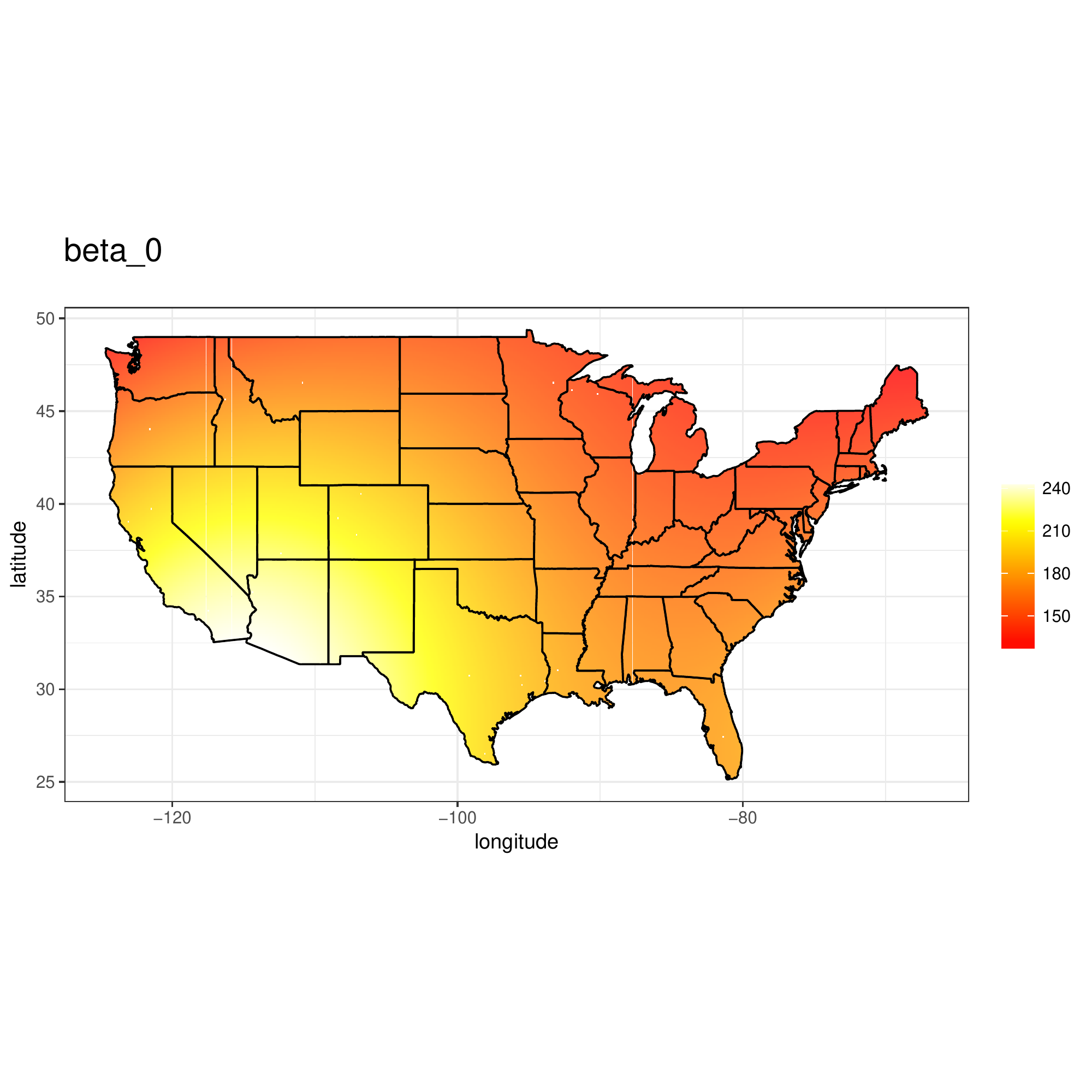} \\
\hfill \includegraphics[scale=0.4,trim=0 120 0 120,clip=TRUE]{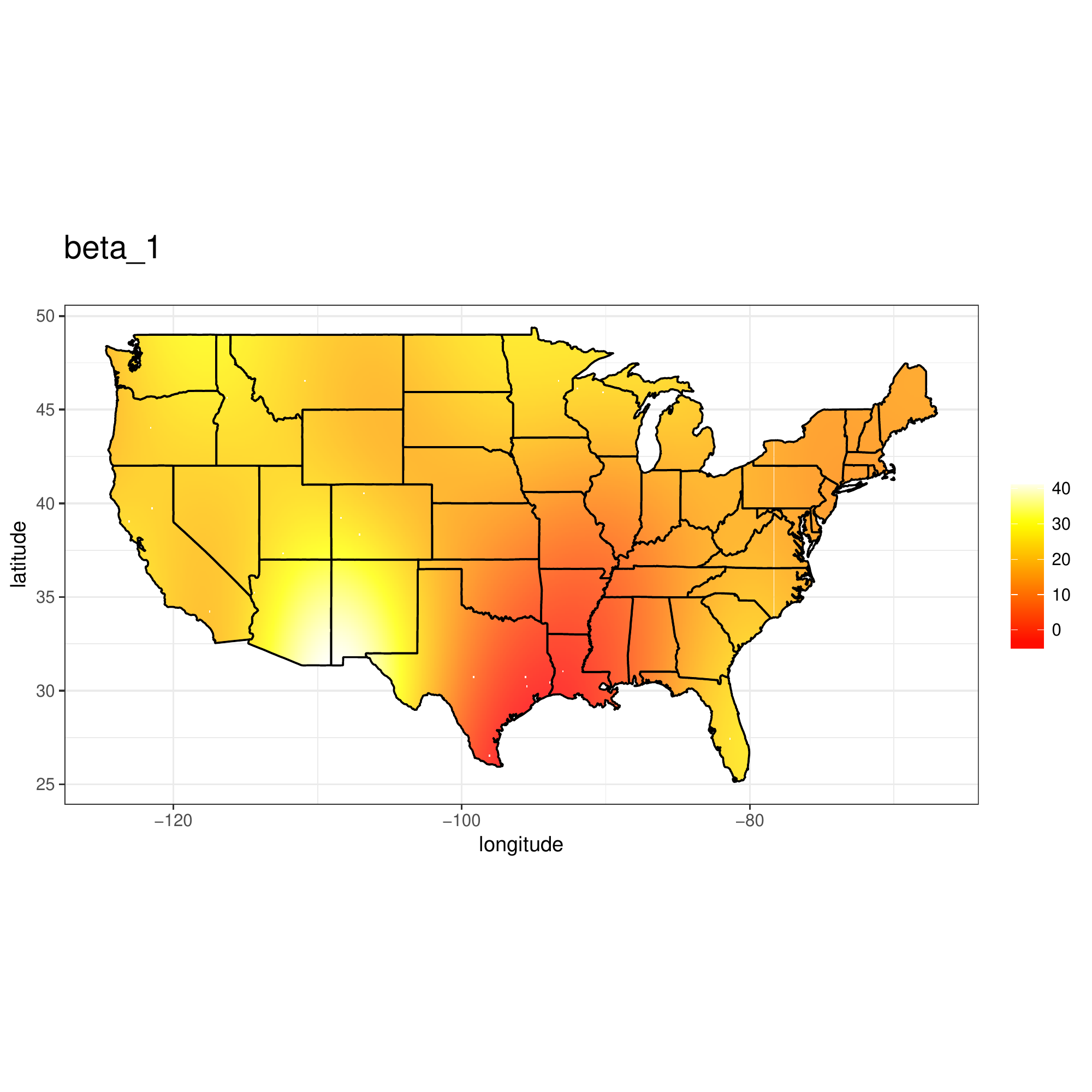} \hfill
\includegraphics[scale=0.4,trim=0 120 0 120,clip=TRUE]{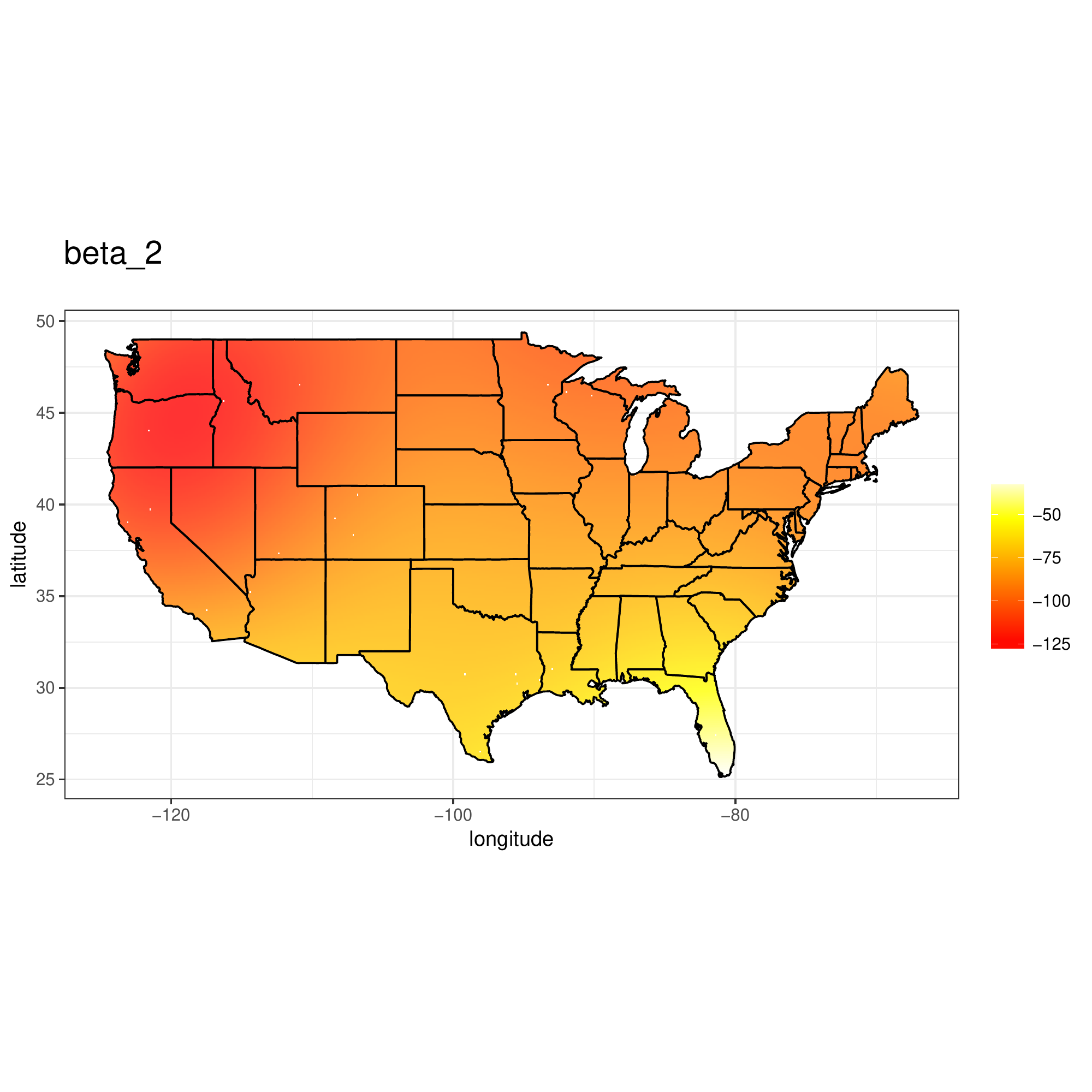}
\caption{The top-left panel shows OLS and smoothed $\beta$-values marginally,
and the remaining three panels show the spatial distribution of the smoothed
$\tilde{\beta}$ values.  Recall that $\beta_0$ is the intercept, $\beta_1$ is
the $\sin$ trigonometric term, and $\beta_2$ the $\cos$.}
\label{f:betas}
\end{figure}

As an illustration, Figure \ref{f:betas} shows $\hat{\beta}$ and
$\tilde{\beta}$ values obtained for our daily field data observations.  The
top-left panel of the figure summarizes the marginal distribution of the two
estimates, which are quite similar except that the distribution of
$\tilde{\beta}$ values is narrower than the un-regularized $\hat{\beta}$s, an
artifact of the regularizing effect of our GP-smoothing. 
The remaining panels
in the figure show the spatial prediction of $\tilde{\beta}(\tilde{s})$ for
$\tilde{s}$ values on our eighty thousand sized predictive grid spanning the
continental US.  Modulo a change in coloring scheme, observe the stark
similarity between the top-right panel, which is for the intercept, and the result
in Figure \ref{f:dtie}.  Seasonal adjustments to this baseline are provided by
the trigonometric periodic coefficients in the bottom panels.  Visualizing
their combination in time is rather more challenging, as that would
effectively involve 365 plots (one for each day of the year).  We delay such a
presentation to Section \ref{sec:vis} after correcting for simulation bias.

\begin{table}[ht!]
\centering \small
\begin{tabular}{rr|r|rrr}
& target &  data/model & RMSE & 95cov & $p$ \\ 
  \hline
1& field& $\widehat{\mathrm{field}}$& 75.86& 0.9594&\\
2& NAM&  $\widehat{\mathrm{NAM}}$& 66.12& 0.9538& \\
3& SREF&  $\widehat{\mathrm{SREF}}$& 73.32& 0.9538& \\
\hline 
4& field&  $\widehat{\mathrm{NAM}}$ w/o b& 79.47& 0.9423& \\
5& field&  $\widehat{\mathrm{SREF}}$ w/o b& 85.70& 0.9405& \\
6& field&  $\widehat{\mathrm{NAM}}$ + $\widehat{\mathrm{b}}$& 75.82& 0.9870&0.03869$^1$\\
7& field&  $\widehat{\mathrm{IVW}}$& 75.81& 0.8908& 4.48e-08$^1$\\
\hline
8& field&  NAM w/o b& 49.82& 0.9680& 0$^7$ \\
9& field&  SREF w/o b& 62.73& 0.9609 \\
10& field&  NAM + b& 44.78& 0.9598& 0$^8$ \\
11& field&  IVW &43.49& 0.8923& $0^{10}$ \\
\end{tabular}
\caption{LOO-CV (average) RMSE output and (average) 95\% predictive coverage
for daily predictors.  See Table \ref{t:loo_cv} for more details.}
\label{t:loo_cv2}
\end{table}

Table \ref{t:loo_cv2} provides a summary of an LOO-CV comparison similar to that
of Table \ref{t:loo_cv}, except in space-time.  Each fold of the CV spans the
entirety of data available in time for that particular spatial location.  When
testing, the accuracy of predictions for all of the available observations in
time (for the particular held-out location in question) is summarized in a
single RMSE value, the spatial average of which is reported in the RMSE column.
Out-of-sample 95\% predictive coverage is reported alongside,
with $p$-values from paired $t$-tests summarizing comparisons up the rows
following in the next column over. The first block in the table provides RMSE
results separately for each of the three targets, $j$. Note that Table
\ref{t:loo_cv2} summarizes a more ambitious prediction exercise, yielding
RMSEs higher than those in Table \ref{t:loo_cv}. However, the range of
observed daily values is also much larger than their time-aggregates, being
from zero to over 1000, so actually in relative terms these forecasts are more
accurate, being at around 4-8\% compared to 11\% previously. Finally,
notice that out-of-sample coverages for these predictors are close to the
nominal level.

\subsection{Bias correction and inverse-variance weighting}
\label{sec:dtie_ivw}

As with the time-aggregated analysis, our ultimate goal is to combine computer
model simulation with field data to obtain a more accurate prediction of
irradiance.  Rows four and five (and nine and ten) show that without
correction for bias, the computer models alone under-impress.  Correcting for
bias in this context involves the same logic as in Section \ref{sec:global},
but in space-time rather than simply space.  Here, to estimate a discrepancy
we form residuals between field data observations and computer model
forecasts, in both space $s$ and time $t$, and train our hierarchical model
(\ref{eq:hier}) on those residuals, ultimately correcting for bias by adding
forecasts obtained for those residuals back into the computer model forecasts
utilized above.  Row six in Table \ref{t:loo_cv2} shows that such corrections
are valuable indeed, albeit on a somewhat smaller scale than in the
time-aggregated case.

Finally, we extend the IVW scheme to the space-time context in order to
combine all three predictors: field, and bias-corrected NAM and SREF.  This
involves extracting the variance of $\tilde{\beta}^j(\tilde{s})$ from the GP
predictive equations (\ref{eq:gppred}) and propagating that through the linear
equation in (\ref{eq:hier}) to derive the variance of the ultimate forecast.
Observe that such a calculation would involve squaring the values of the
trigonometric predictors multiplying the $\beta$-values in the ultimate linear
combination of variances of those components.  In the case of our bias
corrected forecasts, we would need to apply the linear combination twice: once
for the surrogate on NAM or SREF, and again for the bias correction. Observe
in Table \ref{t:loo_cv2} that ``$\widehat{\mathrm{NAM}} +
\widehat{\mathrm{b}}$'' is almost as good as ``$\widehat{\mathrm{IVW}}$''.  In
fact, the latter is not statistically better than the former, which is why we
choose to make the paired $t$-test comparison to the first row for both.
Apparently, SREF (even bias corrected, see below) does not add much value in
this setting, compared to our aggregated results in Table \ref{t:loo_cv}.
Still, from a robustness perspective, we draw comfort from its inclusion in
the ``$\widehat{\mathrm{IVW}}$'' comparator without detriment to accuracy.
The attentive reader will notice [in both Tables \ref{t:loo_cv} and
\ref{t:loo_cv2}] that our IVW comparators under-cover, a point which we
will revisit in our discussion in Section \ref{sec:discuss}.

To conclude the relative comparison portion of this section, we note that in
the unrealistic case where ad hoc computer model simulations are available,
utilizing bias corrected (IVW) versions of those forecasts lead to even better
predictions, effectively establishing a lower baseline for the best possible
results in this context.  We have not included a nonstationary analogue like
{\tt laGP} in the table because we found no improvement from the additional degree
of local modeling this implies.  Observe that our hierarchical model
(\ref{eq:hier}) already facilitates a local-global trade-off, but one
emphasizing time for the former and space for the latter.  The extra
complexity inherent in further localization on the spatial scale (i.e.,
additional estimation risk) would seem to outweigh any benefit that comes from the
added flexibility.  Therefore we choose not to summarize these results in our
table.

\subsection{Visualizing daily forecasts}
\label{sec:vis}

In our supplementary material we provide a movie, as \verb!solar.tar.gz!\footnote{It is available at \url{http://bobby.gramacy.com/solar/solar_irradiance.html} during review.},
compressing 365 days into one minute of elapsed time, showing how irradiance
evolves in space and time, using our best method from Table \ref{t:loo_cv2}
($\widehat{\mathrm{IVW}}$), over the eighty thousand element continental grid.
It is worth reiterating that we don't have the ability to make ad hoc NAM and
SREF evaluations at eighty thousand novel locations $\tilde{s}$, so
comparators from the bottom of Table \ref{t:loo_cv2} are not applicable in
this context.  Most of the movie progression would be confirmatory to someone
experienced in US geography.  
\begin{figure}[ht!]
\centering
\vspace{0.5cm}
\includegraphics[scale=0.4,trim=5 30 50 30,clip=TRUE]{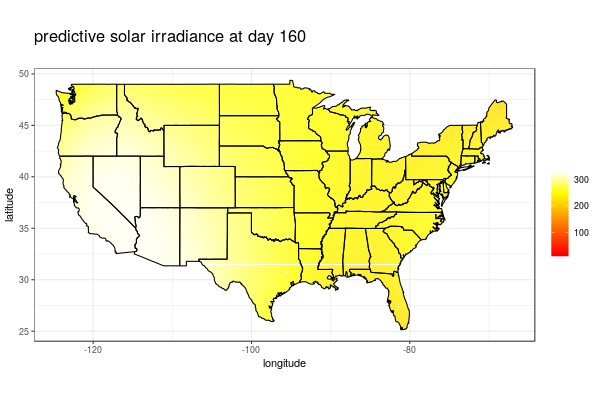} \hfill
\includegraphics[scale=0.4,trim=0 30 10 30]{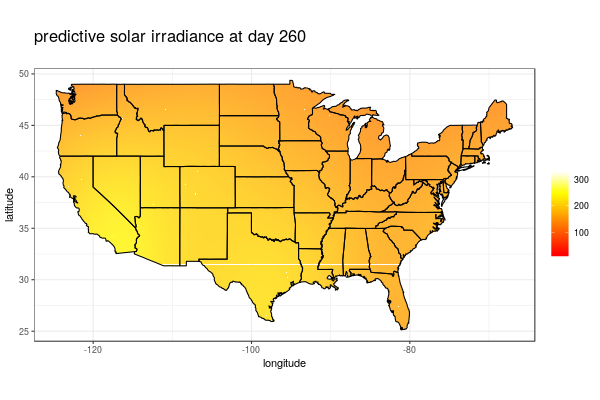}
\caption{Two snapshots of predicted solar irradiance 100 days apart.}
\label{f:jump}
\end{figure}
Figure \ref{f:jump} provides two snapshots spanning a middle 100 days in the
yearly sequence, from day 160 (left panel; June 9) 
to 260 (right panel; September 17). 

\begin{figure}[ht!]
\centering
\includegraphics[scale=0.42,trim=0 110 60 110,clip=TRUE]{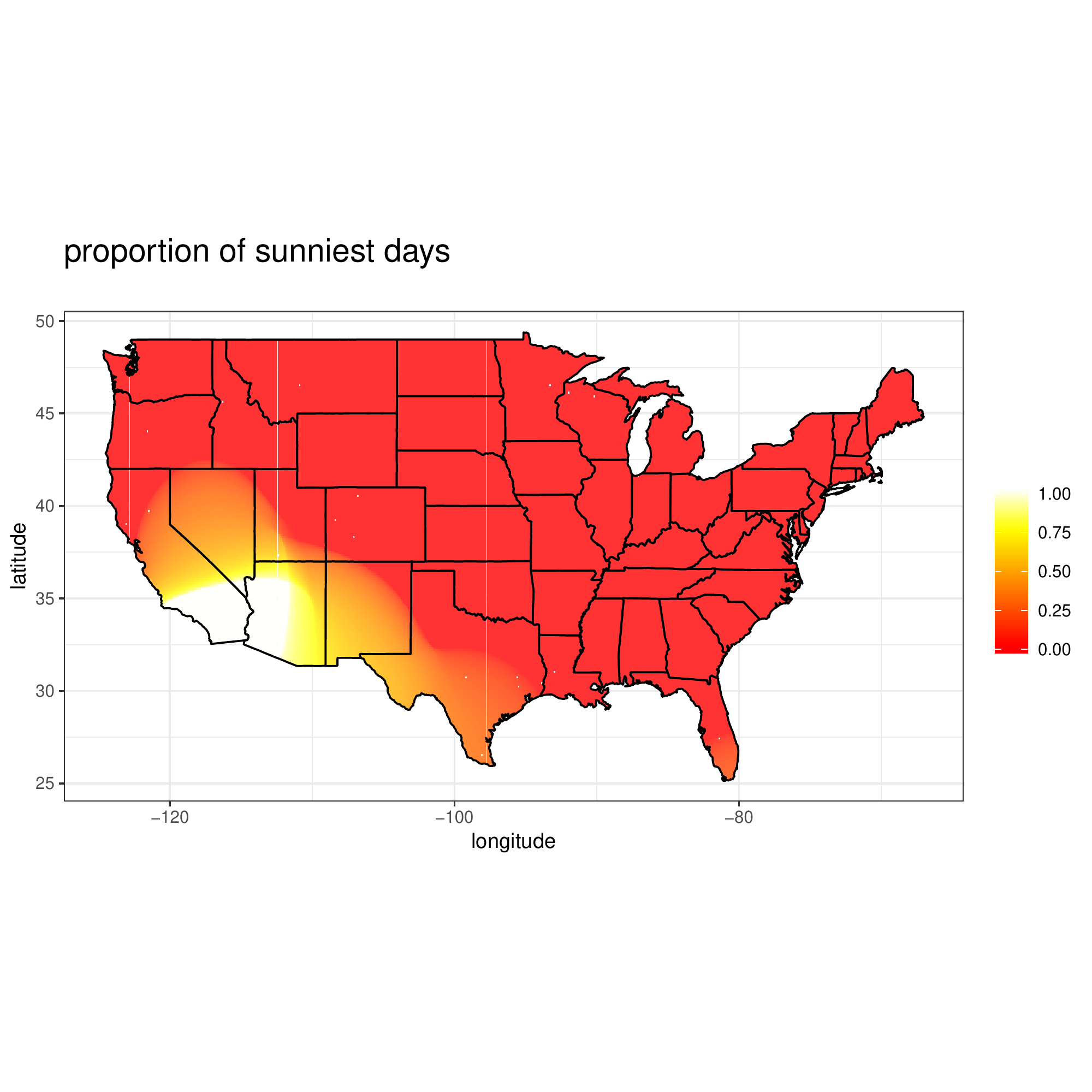} \hfill
\includegraphics[scale=0.42,trim=20 110 10 110]{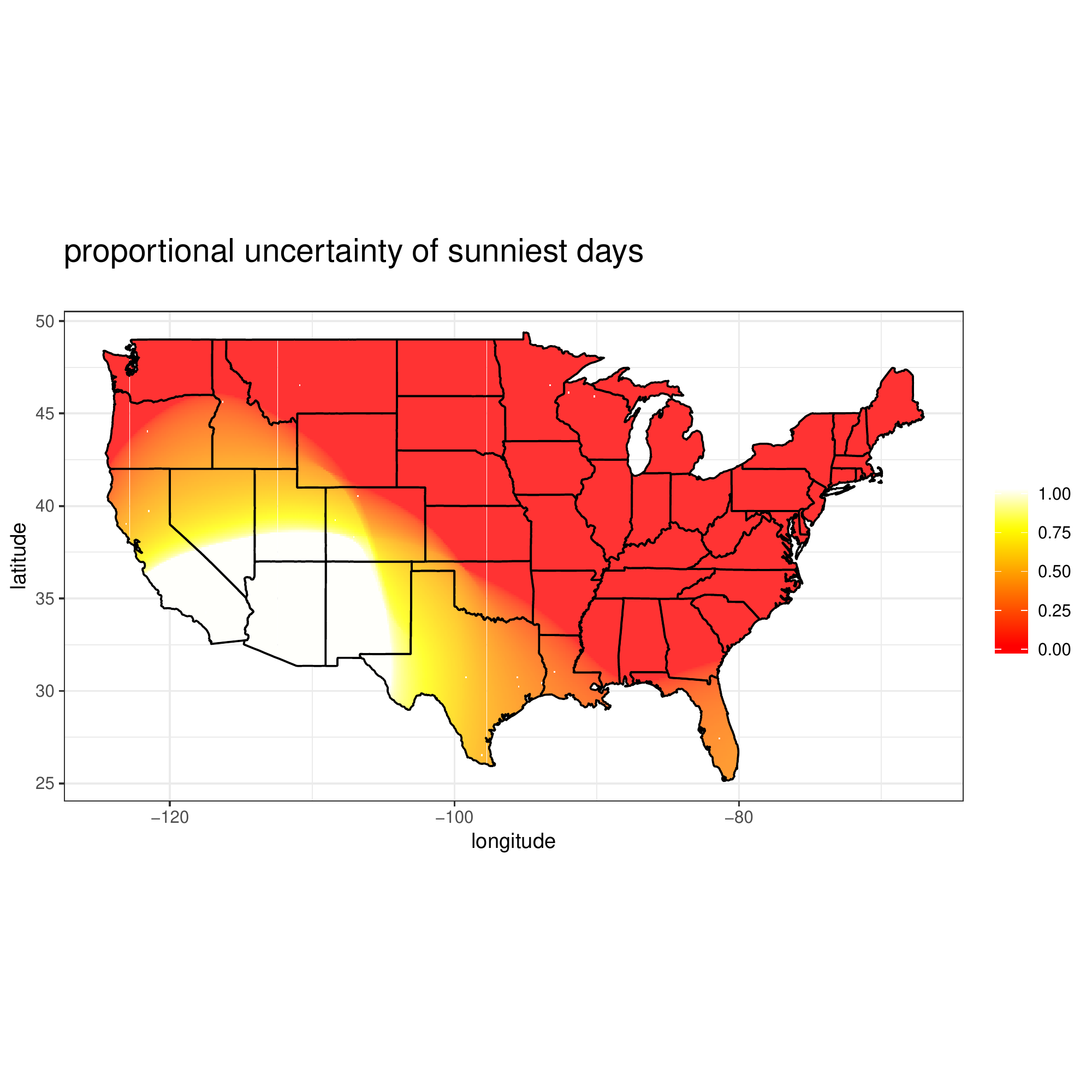}
\caption{The {\em left} panel shows the proportion of time in terms of top 10\% of predictive means, 
and the {\em right} panel shows those in the top quartile, but whose predictive quantile does not 
extend below the bottom quartile of means}
\label{f:sunmean}
\end{figure}

To more compactly summarize the movie we offer the following two additional
views.  Figure \ref{f:sunmean} isolates the most sun-drenched regions in the
USA via heat plots.  The left panel shows the proportion of days where the
particular region is in the top 10\% of the sunniest locations in the USA, in
terms of predicted mean irradiance. It is clear that the Southwest, 
especially Southern California and Arizona, has the highest proportion of the sunniest days. 
As a means of augmenting with predictive uncertainty, the right panel
makes a similar tally, reporting the proportion of time each location had a
predictive mean in the upper quartile (among other predictive means), and its
90\% predictive interval does {\em not} extend below the lower quartile of
that same distribution of means.

\section{New computer model runs}
\label{sec:ip}

The distribution of weather stations [Figure \ref{f:dt_orig}] is clumpy.
Coverage is particularly sparse in regions of the continent where land 
is available for solar farming: in rural parts of the Southwest.  But the
computer models need not be evaluated only at weather stations. Given 
the value of simulations demonstrated in earlier sections, it could be
insightful to determine the extent to which additional runs obtained in 
sparsely sampled parts of the input space could help even more.  Unfortunately 
SREF is not readily available for query, but new NAM runs can be obtained 
from the IBM PAIRS API (\url{https://ibmpairs.mybluemix.net}).  Requests 
may be submitted through their web interface, one geographical coordinate 
at a time (but for a range of days), by selecting ``USA Weather Forecast'' 
in the Dataset category, and ``Solar Irradiance'' in the Datalayer category.  
It is easier to obtain runs for several geographical locations at once through 
{\tt curl}, for example with the following query string for four coordinates on April 14, 2016:

{\singlespacing 
\begin{verbatim}
{"layers":[{"id":"1400"}],"temporal": {"intervals":
  [{"start":"2016-04-14T23:00:00Z","end":"2016-04-15T00:00:00Z"}]},
"spatial":{"type":"point","coordinates":
  37.6642,-121.6073,37.6969,-121.6073,37.6642,-121.5746,37.6969,-121.5746]}}
\end{verbatim} }

The PAIRS database operates on its own grid which has a step size in latitude
and longitude of $1e^{-6}\times2^{15} = 0.032768$ degrees.  If you supply a
coordinate off that grid, you'll get back the response at the nearest grid
location.  Since our weather stations are off that grid, the data we
used were collected from PAIRS NAM runs at the four nearest on-grid locations,
then bilinear interpolation was performed.  The four grid locations in the
query above were the ones used for (lat, lon) $= (37.69, -121.6)$, giving a
final irradiance value of 375.37 after interpolation for that particular
day in April.

\begin{figure}[ht!]
\centering
\includegraphics[scale=0.6,trim=30 160 30 110]{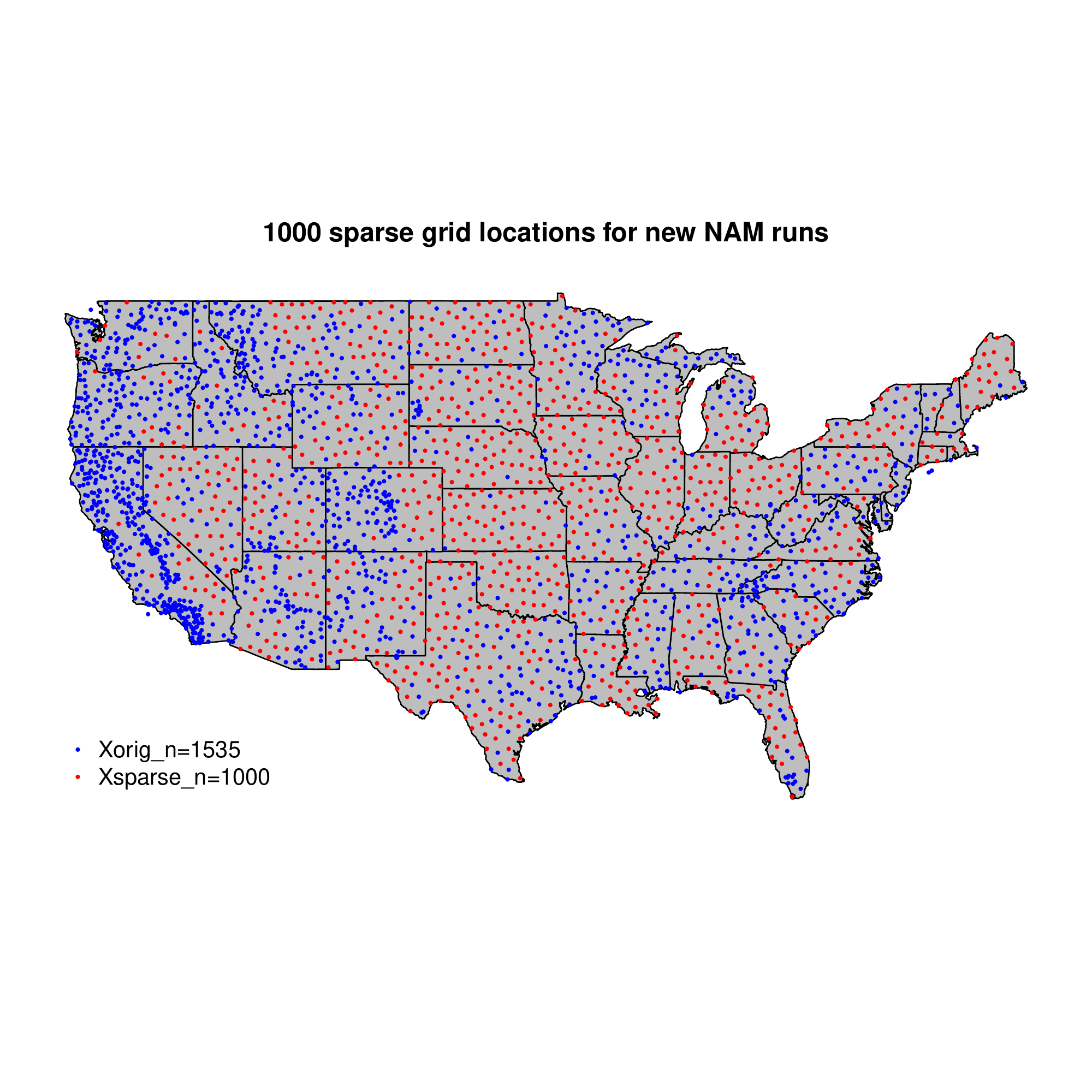}
\caption{1000 space-filling grid locations (red) under a maximin criterion to
themselves and to previous weather station locations (blue).}
\label{f:ip_sparse}
\end{figure}

For our new NAM runs we have some freedom on the precise geographical
location so we figured it would be the most expedient (reducing the number of
evaluations) to choose on-grid locations, avoiding interpolation.  We created
a size 1000 space-filling design by maximizing the minimum distance both to
the existing weather station locations, and between the newly chosen sites.
Then, we snapped those locations onto the NAM/PAIRS grid.  Those locations are
shown as red dots in Figure \ref{f:ip_sparse}, with the original 1535 ones in blue.

Using the augmented cache of NAM runs, now with daily observations at 2513
locations (22 requests in the Northeast, particularly in Maine, were rejected), we repeated 
the LOO-CV experiment in the previous section, which was summarized in Table \ref{t:loo_cv2}.  
The experiment is identical except that the NAM data set is bigger; we are 
still performing LOO-CV over the 1535 weather station locations.
\begin{table}[ht!]
\centering \small
\begin{tabular}{rr|r|rrr|r}
& target &  data/model & RMSE & 95cov & $p$ & $p$-tab\ref{t:loo_cv2}\\ 
\hline
2& NAM &  $\widehat{\mathrm{NAM}}$ &66.18& 0.9508&& 0.003981 \\
\hline 
5& field &  $\widehat{\mathrm{NAM}}$ w/o b& 79.37& 0.9391&& 0.0001907 \\
6 & field &  $\widehat{\mathrm{NAM}}$ + $\widehat{\mathrm{b}}$& 75.79& 0.9865& 0.01354$^1$& 0.08579 \\
7 & field &  $\widehat{\mathrm{IVW}}$& 75.79& 0.8901& 4.803e-11$^1$& 3.635e-05 \\
\hline
\hline
2'& NAM &  $\widehat{\mathrm{NAM}}$ &66.26&  0.9494&& \\
\hline 
5'& field &  $\widehat{\mathrm{NAM}}$ w/o b& 78.85& 0.9374&& 0 \\
6' & field &  $\widehat{\mathrm{NAM}}$ + $\widehat{\mathrm{b}}$& 75.66& 0.9863& 1.821e-07$^1$& 1.596e-06 \\
7' & field &  $\widehat{\mathrm{IVW}}$& 75.73& 0.8895& 0$^1$& 0 \\
\end{tabular}
\caption{LOO-CV (average) RMSE output and (average) 95\% 
out-of-sample predictive coverage for daily predictors on the augmented NAM
dataset with 1000 and 11000 (') new runs.  See Table \ref{t:loo_cv}
for more details.  The $p$ column has the same interpretation as the
$p$-value calculated from a pair-wise one tail $t$-test comparing to the best
method above that row in the table.  The new, final $p$-tab\ref{t:loo_cv2}
column offers a similar comparison to the same row in Table \ref{t:loo_cv2},
showing the benefit of the new NAM runs.}
\label{t:loo_cv3}
\end{table}
The results of this experiment are shown in Table \ref{t:loo_cv3}.  The
interpretation of the columns is the same as in earlier tables, with the
exception of a new ``$p$-tab\ref{t:loo_cv2}'' column.  Note that this table
has fewer rows---the missing ones would be identical to ones in Table
\ref{t:loo_cv2}. The new ``$p$-tab\ref{t:loo_cv2}'' column offers a comparison
to the same row of Table \ref{t:loo_cv2}, quantifying the benefit of our new
NAM runs in statistical terms.

Our analysis of the results of this experiment are nuanced.  It is clear that 
$\widehat{\mathrm{NAM}}$ surrogate accuracy is improved, not by a large margin 
but by a significant one.  Predictors that are based directly on that surrogate, 
without adjusting for bias, also saw significant improvement over our earlier 
results.  Those which adjust for that bias, however, saw improvement but not 
at a level that is statistically significant.  In particular, the 
``$\widehat{\mathrm{NAM}}+\widehat{\mathrm{b}}$'' comparator is not better 
than the previous one.  $\widehat{\mathrm{IVW}}$ is better this time around, 
but we find that hard to explain when the only part of that comparator 
which changed is that to do with its ``$\widehat{\mathrm{NAM}}+\widehat{\mathrm{b}}$''.  
We conjecture that the improvements in the $\widehat{\mathrm{NAM}}$ surrogate, 
particularly a reduced predictive variance owing to the larger training dataset, 
led to a more advantageous weight assignment in the IVW scheme.

Upon request from a referee, we have obtained NAM simulations 
on an additional set of space-filling locations from IBM PAIRS; out of $10000$
further requests we received $9698$ responses. The $302$ missing ones are
again from the Northeast.  With a training dataset of this size, augmenting
the ones above (for a total of $12211$ locations), only a local GP surrogate
is computationally feasible.  LOO-CV obtained with that surrogate, but otherwise
with an identical setup to the one described above, are provided in the
augmented Table \ref{t:loo_cv3}. Compared to the first set of space-filling
locations, the emulator based on more training data has better predictive
performance, which makes sense. For example consider rows 6 and 6'
corresponding to the sum of emulated NAM and its estimated discrepancy via 1000
and 10000 extra NAM runs, respectively. Row 6 indicates predictive performance
is not significantly better than its counterpart in Table \ref{t:loo_cv2}; row
6', however, is statistically superior.

\section{Discussion}
\label{sec:discuss}

We set up a statistical framework to analyze a suite of data combining
geographic measurements and two computer model outputs of solar irradiance 
at 1535 spatial locations across the continental United States. Perhaps 
the most important takeaway result is that the computer model, when 
suitably bias-corrected, offers a quite accurate forecast of observed irradiance.
This is true whether using available simulations, or emulating those
simulations at geographical locations where none are readily available.  The
former is perhaps more realistic for most applications of such an analysis,
e.g., geo-locating future solar farms.

We showed that a local--global tradeoff offered the best results in terms of
spatial and spatial--temporal modeling (and bias correction).  In the case of
time-aggregated observations, we found that a local approximate Gaussian
process ({\tt laGP}) could better cope with the differential dynamics at play
in disparate geographic regions, say the Southwest versus Northeast.  Our
model for the daily data involved local time series smoothed over space,
facilitating an organic local-global partition. We did not find any benefit to
further spatial localization in this context. Finally, we showed that extra
computer model runs, when available, could be used to obtain slightly more
accurate predictions in time and space. Our analysis only considered
a yearly cycle, but this could easily be extended with additional data (we
only had 1.5 years).  For example, with decades of data we could entertain 
the effects of an 11-year magnetic-pole cycle.

One point worth mentioning is that predictive accuracy should not 
be the only focus in forecasting.   Incorporating predictive variability is
essential, and our method could be improved on that front. For example, our
IVW scheme provides the best predictive accuracy, but the worst predictive
coverage compared to nominal.  One explanation might be that correlation
exists between the predictors being combined, which means that our weighted
average of variances is under-estimating.  In future work we plan to augment
the IVW scheme with an explicit estimate of out-of-sample covariance in order
to make the requisite adjustments.

Short term accuracy of suitably bias-corrected weather model-based fine-scale
simulation of irradiance, when realizations are available in real space-time
(e.g., in future days), is potentially viable but remains illusive when
(future) weather is unavailable to be conditioned upon.  In more realistic
contexts, smoothed dynamics offered by (bias corrected) surrogates offer a
surprisingly good alternative. This mirrors recent results in similar contexts
where nonparametric models are deployed to ``memorize'' patterns in the data
at the expense of learning precise (even strong) correlations between events.
For example, \citet{johnson:etal:2018} show that GPs directly on dengue
incidence dynamics, ignoring strong temporal correlation between covariates
like precipitation and transmission, provide more accurate predictions than
precipitation-based models.  The explanation is that forecasting precipitation
accurately is harder than the original (dengue-tracking) problem. In our
context, although it is known that cloud cover effects irradiance, it is hard
to predict cloudiness on a particular day in a typical year.  Yet we showed
that doesn't mean that forecasts based on weather models are useless.
Surrogates can be deployed to extract seasonality and large-scale spatial
variability from the models to improve upon the accuracy obtained via fitting
to the field data alone (i.e., via a more purely machine-learning approach), 
so long as bias can be suitably corrected.
All of the code and data supporting our empirical work is available in a public
version-controlled repository:
\url{https://bitbucket.org/gramacylab/solance}.  The source files for the
experiments may be found in the \verb!R! directory, which reference
(text/csv) data files provided in
\verb!data!.  This completely self-contained implementation, modulo
the {\sf R} packages which are loaded by the code, supports a fully
reproducible analysis for all tables and figures presented in this manuscript.

We'd like to close by featuring a potential avenue for further research.  We
chose a thrifty alternative to fully Bayesian inference for our hierarchical
space-time model in Eq.~(\ref{eq:hier}).  Assuming a tractable MCMC scheme
could be developed for fully Bayesian inference---which would be challenging
considering the high dimensionality (in the thousands) 	of the latent $\beta$
variables---it would be interesting to investigate how, or whether, such
diligence leads to different results. Fuller assessments of the underlying
uncertainties, a hallmark of Bayesian inference, could lead to more accurate
selections of low volatility/high irradiance regions, such as those identified
in the panels of Figure \ref{f:sunmean}.  However, this comes at the
risk of difficult-to-verify MCMC convergence in such high dimensions.

\subsection*{Acknowledgments}

The authors would like to gratefully acknowledge funding from the National 
Science Foundation, awards DMS-1564438, DMS-1621722, DMS-1621746, and 
DMS-1739097, as well as the resources provided by Advanced Research Computing 
at Virginia Tech.

\appendix

\bigskip

\section{GP versus BART}
\label{sec:bart}

We are grateful to a referee for suggesting that we compare our GP-based
predictors with Bayesian additive regression trees
\citep[BART,][]{chipman:2010} via {\tt BayesTree} \citep{bayestree} on CRAN.
That comparison is most straightforward in the context of our time-aggregated
10-fold CV in Section \ref{sec:ta}, via RMSE.  However, a big difference
between BART and GP-based methods is in the form of the predictive variance.
Therefore for the comparison here we augment RMSE with a proper scoring rule
in \citet{gneiting:raftery:2007}: $S(\hat{y},y) = -\log |\Sigma_{\hat{y}}| -
\left(y-\hat{y}\right)^\top\Sigma^{-1}_{\hat{y}}\left(y-\hat{y}\right)$.
Higher scores are better.
\begin{figure}[ht!]
\centering
\includegraphics[scale=0.32,trim=0 30 0 40]{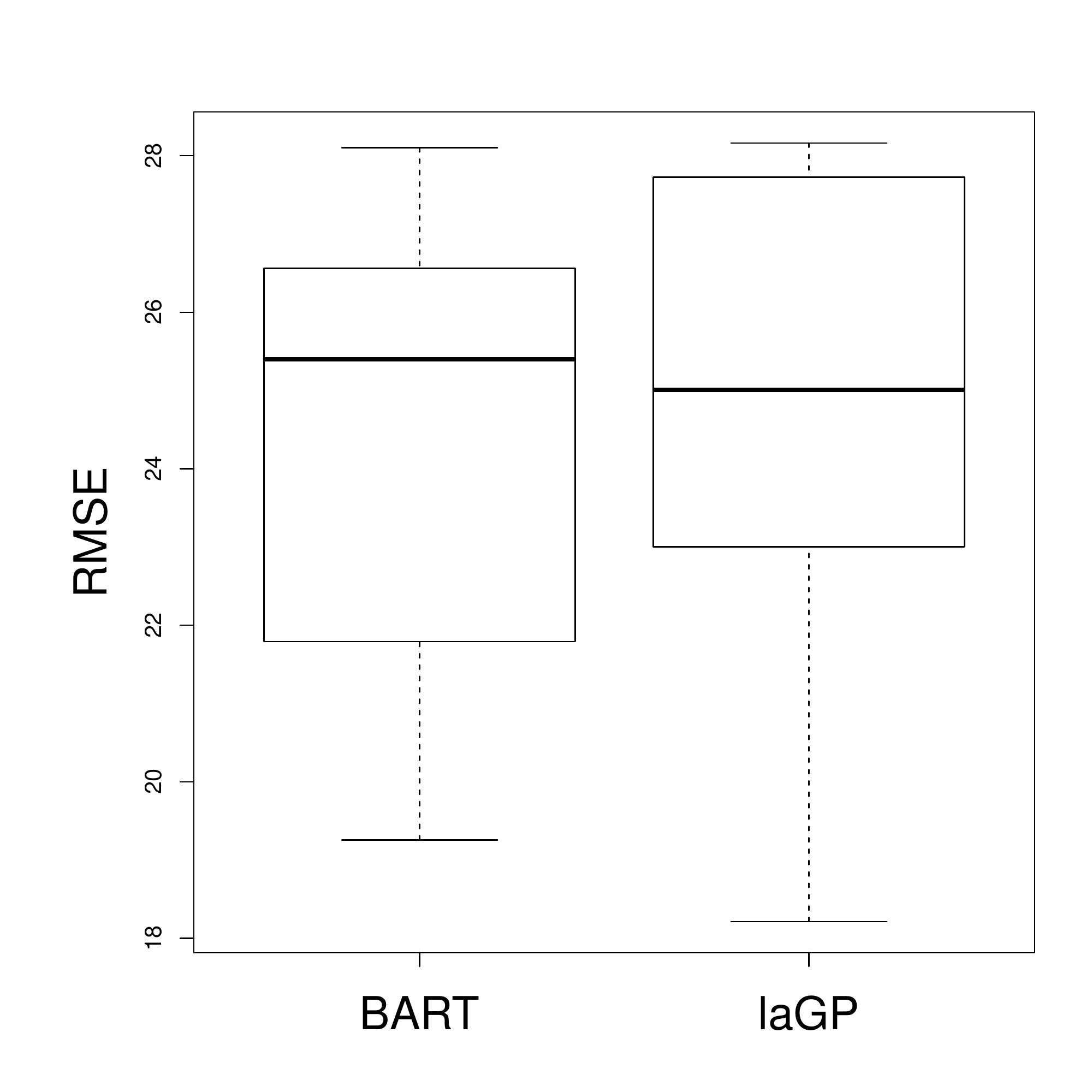}
\includegraphics[scale=0.32,trim=0 30 0 40]{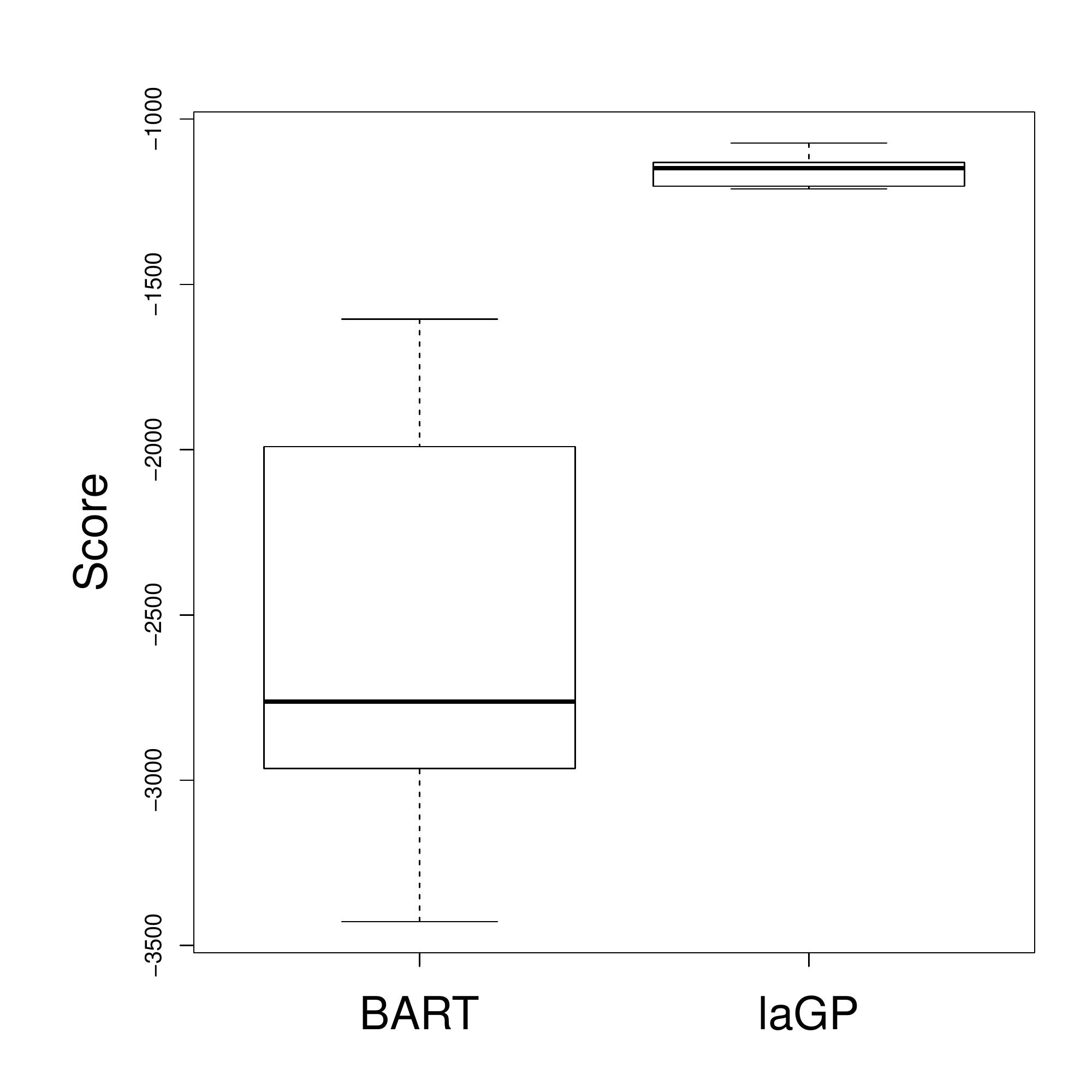}
\caption{Comparison in terms of RMSE ({\em left}) and proper score ({\em
right}) between BART and {\tt laGP} on time-aggregated field measurements 
in the framework of 10-fold cross-validation.}
\label{f:bart}
\end{figure}
These results are summarized in Figure
\ref{f:bart}.  Observe that BART is comparable to {\tt laGP}, which is the
same {\tt laGP} comparator from Section \ref{sec:ta}, on RMSE grounds 
but inferior in terms of proper score.  The homoskedastic additive variance
assumption in BART is inferior in this context.

\newpage
\bibliography{../laGP/laGP,../gpu/gpu,../rays/rays,../calib/calib,../satdrag/satdrag,solance,rebuttal}
\bibliographystyle{unsrtnat}

\end{document}